\documentclass[12pt]{article}
\usepackage{epsfig,cite}
\usepackage {amsmath}
\usepackage{amssymb}
\usepackage{color}
\include{epsf}
\newcount\timecount
\newcount\hours \newcount\minutes  \newcount\temp \newcount\pmhours
\hours = \time
\divide\hours by 60
\temp = \hours
\multiply\temp by 60
\minutes = \time
\advance\minutes by -\temp
\def\hour{\the\hours}
\def\minute{\ifnum\minutes<10 0\the\minutes
            \else\the\minutes\fi}
\def\clock{
\ifnum\hours=0 12:\minute\ AM
\else\ifnum\hours<12 \hour:\minute\ AM
      \else\ifnum\hours=12 12:\minute\ PM
            \else\ifnum\hours>12
                 \pmhours=\hours
                 \advance\pmhours by -12
                 \the\pmhours:\minute\ PM
                 \fi
            \fi
      \fi
\fi
}

\def\monthname{\relax\ifcase\month 0/\or January\or February\or
   March\or April\or May\or June\or July\or August\or September\or
   October\or November\or December\else\number\month/\fi}

\def\bold#1{\setbox0=\hbox{$#1$}%
     \kern-.025em\copy0\kern-\wd0
     \kern.05em\copy0\kern-\wd0
     \kern-.025em\raise.0433em\box0 }

\def\beq{\begin{equation}}
\def\eeq{\end{equation}}

\def\ga{\mathrel{\raise.3ex\hbox{$>$\kern-.75em\lower1ex\hbox{$\sim$}}}}
\def\la{\mathrel{\raise.3ex\hbox{$<$\kern-.75em\lower1ex\hbox{$\sim$}}}}
\def\gev{{\rm \, Ge\kern-0.125em V}}
\def\tev{{\rm \, Te\kern-0.125em V}}
\def\gyr{{\rm \, G\kern-0.125em yr}}

\def\gappeq{\mathrel{\rlap {\raise.5ex\hbox{$>$}}
{\lower.5ex\hbox{$\sim$}}}}
\def\lappeq{\mathrel{\rlap{\raise.5ex\hbox{$<$}}
{\lower.5ex\hbox{$\sim$}}}}
\def\Toprel#1\over#2{\mathrel{\mathop{#2}\limits^{#1}}}

\def\m12{m_{1\!/2}}

\def\bea{\begin{eqnarray}}
\def\eea{\end{eqnarray}}

\def\beq{\begin{equation}}
\def\eeq{\end{equation}}

\begin{document}

\begin{titlepage}
\pagestyle{empty}
\baselineskip=21pt
\rightline{UMN--TH--3207/13, FTPI--MINN--13/18, IPMU13-0116}
\vspace{0.2cm}
\begin{center}
{\large {\bf Non-Universalities in Pure Gravity Mediation }}
\end{center}
\vspace{0.5cm}
\begin{center}
{\bf Jason L. Evans}$^{1}$,
{\bf Masahiro Ibe}$^{2,3}$ {\bf Keith~A.~Olive}$^{1}$
and {\bf Tsutomu T. Yanagida}$^{3}$\\
\vskip 0.2in
{\small {\it
$^1${William I. Fine Theoretical Physics Institute, School of Physics and Astronomy},\\
{University of Minnesota, Minneapolis, MN 55455,\,USA}\\
$^2${ ICRR, University of Tokyo, Kashiwa 277-8582, Japan}\\
$^3${Kavli IPMU (WPI), TODIAS, University of Tokyo, Kashiwa 277-8583, Japan}\\
}}
\vspace{1cm}
{\bf Abstract}
\end{center}
\baselineskip=18pt \noindent
{\small The simplest model of pure gravity mediation contains only two free parameters:
the gravitino mass and $\tan \beta$. Scalar masses are universal at some high
energy renormalization scale and gaugino masses are determined through anomalies
and depend on the gravitino mass and the gauge couplings.  This theory requires
a relatively large gravitino mass ($m_{3/2}\gtrsim 300$\,TeV)
and a limited range in $\tan \beta \simeq 1.7$--$2.5$.
Here we generalize the theory to allow for non-universality in the Higgs soft masses.
This introduces zero, one or two new free parameters associated with Higgs soft masses.
This generalization allows us to greatly increase the allowed range in $\tan \beta$
and it allows one to find viable solutions with lower $m_{3/2}$. The latter
is important if we hope to find a low energy signal from gluinos. Some special cases of these
non-universalities are suggestive of Higgs bosons as Nambu-Goldstone bosons or
a partial no-scale structure for the Higgs doublets. Thus, we probe signatures at the weak
scale and structures at the GUT and/or Planck scale.
}


\vfill
\leftline{June 2013}
\end{titlepage}

\section{Introduction}

Recent results at the LHC indicate a Higgs mass of roughly 126 GeV \cite{lhch}
which is near the upper end of the range allowed in simple and commonly
studied models such as the constrained minimal supersymmetric standard model (CMSSM) \cite{cmssm} and indeed has put pressure on these models pushing the mass scales
associated with supersymmetry breaking parameters to higher values \cite{125-other}.
Furthermore, the  absence
of supersymmetric particles at the LHC~\cite{lhc} also point to higher mass scales \cite{postlhc}.
While successful phenomenologies can still be constructed for the CMSSM \cite{eo6},
the data seems to point beyond the CMSSM \cite{elos}.

Recently, we have shown \cite{eioy} that models based on pure gravity mediation (PGM)
\cite{pgm,pgm2,ArkaniHamed:2012gw} with full scalar mass universality are consistent
with existing experimental constraints.  These models contain only two free continuous parameters:
the gravitino mass $m_{3/2}$ and the ratio of the two Higgs vevs, $\tan \beta$
(the sign of the Higgs mixing mass, $\mu$ must also be specified).
Scalar masses are universal and equal to the gravitino mass at some high energy input scale,
usually assumed to be the GUT scale
(the renormalization scale where the electroweak gauge couplings are equal).
The two additional parameters that define the CMSSM, a universal gaugino mass and a
universal tri-linear coupling are generated through anomalies \cite{anom} in PGM.
As a consequence, $m_{1/2}, A_0 \ll m_0$ in these models -- a common characteristic of
models of split supersymmetry \cite{split}.  Similar models have been shown to arise
when moduli (volume moduli and those associated with supersymmetry breaking)
are strongly stabilized \cite{klor,dlmmo,Dudas:2006gr}.

Because gaugino masses are loop suppressed, the LEP limit on the chargino mass
\cite{LEPsusy} pushes the gravitino mass to at least 30 TeV and at low $\tan \beta$,
the Higgs mass generally pushes $m_{3/2}$ even higher.
With such high values of the universal scalar mass, it is not obvious that
one can construct a viable theory and maintain one of the most desirable features
of CMSSM-like models, namely radiative electroweak symmetry breaking \cite{ewsb}.
It is indeed possible to construct such models \cite{eioy}
if $\tan \beta$ is restricted to the relatively narrow range between $1.7$--$2.5$.
For $m_{3/2} \simeq 300$--$1500$ TeV, one can obtain $m_h \simeq 126 \pm 2$ GeV.
At such large mass scales, the only supersymmetric observable may
be a relatively long lived chargino which is nearly degenerate with the neutral
SU(2) gaugino \cite{pgm,pgm2,ggw,Feng:1999fu,ctchar,dlmmo}. Furthermore, it was shown that for $\mu < 0$, thermal (wino)
dark matter with a relic density equal to the WMAP/Planck \cite{wmap} determined value
is expected when $m_{3/2} = 460$--$500$\,TeV. For lower $m_{3/2}$,
either the dark matter comes from a source other than supersymmetry,
or winos are produced non-thermally through moduli or gravitino
decay \cite{ggw,hep-ph/9906527,Ibe:2004tg,Acharya:2008bk,dlmmo}.

In \cite{eioy}, we considered only models with full scalar mass universality.
Here we generalize these PGM models to allow for Higgs mass non-universality
as in the non-universal Higgs mass (NUHM) models \cite{nonu,nuhm2,nuhm1}.
Non-universality in the Higgs sector is well-motivated and is expected in Grand Unified
Theories (GUTs) where the Higgs multiplets live in separate representations from matter fields.
This may result in one additional soft mass parameter (NUHM1 \cite{nuhm1}) if
the two electroweak doublets originate in a common GUT multiplet as in the case
of an SO(10) GUT or two additional parameters (NUHM2 \cite{nuhm2}) if the
electroweak doublets originate in separate multiplets.

Special cases of interest are cases where one or both of the Higgs multiplets
can be associated with pseudo Nambu-Goldstone bosons \cite{Kugo:1983ai}, where
the soft masses of the Higgs doublets are quite suppressed compared to $m_{3/2}$.
This hypothesis not only provides non-universality to the Higgs soft masses
but may also shed some light on the fundamental question:
why are the Yukawa and Gauge couplings perturbatively  small
at the GUT or the Planck scale.
In this paper, we discuss some examples of the models in which the Higgs doublets
are fully or partially realized as pseudo Nambu-Goldstone supermultiplets.
As we will show the highly suppressed Higgs soft masses are consistent with
electroweak symmetry breaking, and hence, the Nambu-Goldstone hypothesis
is viable.

Allowing some non-universality has two direct consequences on
low energy phenomenology: 1) the range of allowed values of $\tan \beta$
is greatly relaxed; 2) the lower limit of $\sim 300$ TeV on the gravitino
mass is relaxed allowing values which drop to the LEP imposed bound of 30 TeV.
At larger $\tan \beta$, the importance of
radiative corrections to neutralino and chargino masses \cite{piercepapa} is greatly diminished
leading to larger (smaller) gaugino/chargino masses when $\mu > 0$ ($< 0$)
reducing the sensitivity to the sign of $\mu$.
The ability to define models with lower $m_{3/2}$
allows for the possibility that viable models can be constructed with
a reasonably light gluino, potentially detectable at the LHC.

In section 2, we define our PGM model which allows for
non-universality in the Higgs sector. In section 3, we concentrate on
NUHM1 like models where the two Higgs soft masses are equal.
In section 4, PGM models are further generalized to the NUHM2 where
both Higgs soft masses are taken as free parameters. In section 5,
we consider the special cases where one or both of the
Higgs doublets originate as Nambu-Goldstone bosons.
Our conclusions are given in section 6.

\section{PGM and Non-Universality\label{PGMnU}}

The construction of universal PGM models was described in
detail in \cite{eioy}. Here, we will generalize the model to allow
for non-universal soft masses in the Higgs sector.

Our starting point is a flat K\"ahler manifold for all Standard Model
fields except the Higgs doublets.
Let us define the following fields:
$H_{1,2}$ the MSSM Higgs doublets (sometimes referred to as $H_{d,u}$ respectively);
$Z$ a supersymmetry breaking field,
and $y$ a generic MSSM scalar field other than the Higgs doublets.
The K\"ahler potential can be written in the form:
\beq
K =   y y^*  + K^{(H)} +  K^{(Z)}  +  \log |W|^2\ ,
\label{K1}
\eeq
where
\beq
K^{(Z)} = Z Z^* \left(1-\frac{Z Z^*}{\Lambda^2}\right) \ ,
\label{kz}
\eeq
and
\beq
K^{(H)} = \left(1 + a \frac{Z Z^*}{M_P^2}\right) H_1 H_1^*
+ \left(1 + b \frac{Z Z^*}{M_P^2}\right) H_2 H_2^* + (c_H H_1 H_2 + h.c.) \, .
\label{kh}
\eeq
Here, $M_P$ denotes the Planck scale, while $\Lambda$ is the typical scale of dynamical SUSY breaking
in which $Z$ takes part.
We assume $\Lambda \ll M_P$ so that the  $Z$ is strongly stabilized \cite{dine,dlmmo}.
Our results here are not sensitive to the particular assumptions on $K^{(Z)}$ or the
value of $\Lambda$.
In Eq.\,(\ref{kh}), we have introduced two new parameters.
The couplings $a,b$ to specify the degree of non-universality in the Higgs sector.
For universal models, we have $ a = b = 0$.
The third coupling, $c_H$ was already present in the universal PGM model of \cite{eioy}
and is a generalized Giudice-Masiero (GM) term \cite{gm,Inoue:1991rk}
needed to be able to simultaneously
choose $\tan \beta$ as a free parameter and still obtain consistent
electroweak symmetry breaking conditions and satisfy all supergravity boundary conditions \cite{dmmo,eioy}.

If we choose a superpotential
\beq
W = M^2 Z + \Delta + W_{MSSM}
\label{W}
\eeq
where $\Delta$ is a constant and $W_{MSSM}$ is the MSSM superpotential which
includes the $\mu$-term, $\mu H_1 H_2$, we arrive at the following
\begin{itemize}
\item A Minkowski vacuum implies $M^4 = 3\displaystyle{\frac{ \Delta^2}{M_P^2}}$.
\item $\langle Z \rangle \simeq \displaystyle{\frac{\sqrt{3} \Lambda^2}{6M_P}}$.
\item $m_Z^2 \simeq  \displaystyle{\frac{12 M_P^2 m_{3/2}^2}{ \Lambda^2}}  \gg m_{3/2}^2$.
\item $m_{3/2} =\displaystyle{ \frac{\Delta}{M_P^2}}$,
and $m_y^2 = m_{3/2}^2$ for all MSSM fields other than Higgs doublets
\item $m_1^2 = (1-3a)m_{3/2}^2$ and $m_2^2 = (1-3b)m_{3/2}^2$.
Here the slight effect due to the non-canonical kinetic term of order ${\cal O}( \Lambda^4/M_P^4)$
is neglected.
\end{itemize}
By choosing $a,b \ne 0$, we generate non-Universal soft masses for
$H_1$ and $H_2$.
For  the special case of $a,b = 1/3$, we have $m_{1,2}^2 = 0$ which we will consider later in
section 5.
In addition, $A$-terms are small, $A_0 \sim (\Lambda^2/M_P^2) m_{3/2} \ll m_{3/2}$,
and thus the dominant contribution to the $A$-terms are generated through anomalies.
If $Z$ carriers an $R$-charge, gaugino masses are also suppressed and the dominant
contributions also come from anomalies \cite{anom}
\begin{eqnarray}
    M_{1} &=&
    \frac{33}{5} \frac{g_{1}^{2}}{16 \pi^{2}}
    m_{3/2}\ ,
    \label{eq:M1} \\
    M_{2} &=&
    \frac{g_{2}^{2}}{16 \pi^{2}} m_{3/2}  \ ,
        \label{eq:M2}     \\
    M_{3} &=&  -3 \frac{g_3^2}{16\pi^2} m_{3/2}\ .
    \label{eq:M3}
\end{eqnarray}
Here, the subscripts of $M_a$, $(a=1,2,3)$, correspond to the gauge groups of the Standard Model,
U(1)$_Y$, SU(2) and SU(3), respectively.

Alternatively, we can envision a (partial) no-scale structure \cite{Cremmer:1983bf} in the
Higgs K\"ahler manifold. Suppose instead of (\ref{K1}), we choose
\beq
K =   y y^* - 3 \log \left( 1 - \frac{1}{3}(H_1 H_1^* + H_2 H_2^*+ K^{(Z)}) \right)
 +  (c_H H_1 H_2 + h.c.)  +  \log |W|^2
\label{K2}
\eeq
with $K^{(Z)} $ and $W$ as in Eqs.\,(\ref{kz}) and (\ref{W}), respectively.
The resulting soft masses for the Higgs doublets in this case is  $m_1^2 = m_2^2 = 0$.
One could also consider
only one of the two Higgs doublets having the no-scale structure.
Of particular interest is the resulting $B$-term. In pure no-scale models,
we expect both $A_0 = 0$ and $B_0 = 0$. Due to the presence of the Higgs bi-linear term,
the no-scale structure is broken and we are able to generate $B(M_{in})  \ne 0$\footnote{Because of the structure of the hidden sector contribution to the K\"ahler potential, the $F$-term of the conformal compensator is non-zero and so we have an anomaly mediated contribution to the soft masses.}.
For a detailed discussion of the Higgs soft mass terms and the $\mu$ and $B\mu$ term
see the appendix \ref{Append}.
The take away lesson is that if the Higgs term $|H_i|^2$ is inside the logarithm,
it has a vanishing soft mass otherwise it is $m_{3/2}$, while
$\mu$ and $B\mu$ are unaffected by the no-scale structure of the Higgs K\"ahler potential.
They are affected by the no-scale structure of the GM term.
However, since we only care about the low scale values of $B\mu$ and $\mu$,
their exact forms are unimportant.
For the K\"ahler potential in Eq.\,(\ref{K2}), the relations are\footnote{In the case of strong moduli stabilization, the $\mu_0$ listed here is really $\mu'$ and is related to the value of  $\mu_0$ in the superpotential by $\mu_0=(\rho+\bar\rho)^{3/2}\mu'$ where $\rho$ is a volume modulus field
(see appendix).}
\begin{eqnarray}
&&B\mu= 2m_{3/2}^2c_H -m_{3/2}\mu_0,\\
&&\mu=c_H m_{3/2} +\mu_0,
\end{eqnarray}
where $\mu_0$ is the bilinear mass term in the superpotential and the Higgs soft masses are zero.

In summary, while the minimal model of PGM  has
only two fundamental parameters: $m_{3/2}$, and the generalized Giudice-Masiero term, $c_H$.
A straightforward generalization of this model adds one or two new parameters
characterizing the non-universality of the Higgs masses.
At the level of the K\"ahler potential, these may be chosen to be the couplings
 $a$ and $b$ in Eq.\,(\ref{kh}).
Relating these to the Higgs soft masses, $m_{1,2}$ and relating $c_H$ to $\tan \beta$
through the electroweak symmetry breaking boundary conditions, our
non-universal PGM model is described by
\begin{eqnarray}
 m_{3/2}\ , \quad m_1\ , \quad m_2\ , \quad \tan\beta \ ,
\end{eqnarray}
or equivalently
\begin{eqnarray}
 m_{3/2}\ , \quad \mu\ , \quad m_A\ , \quad \tan\beta \ .
\end{eqnarray}
Note that in the above lists, we treat $m_1$ and $m_2$ as GUT scale parameters and
$\mu$ and $m_A$ as weak scale parameters.
We will use both parameterization below.
\section{NUHM1 models}

We begin our probe of non-universal PGM models with
the three parameter version of the NUHM with $m_1 = m_2$.

\begin{figure}[ht]
\begin{minipage}{8in}
\epsfig{file=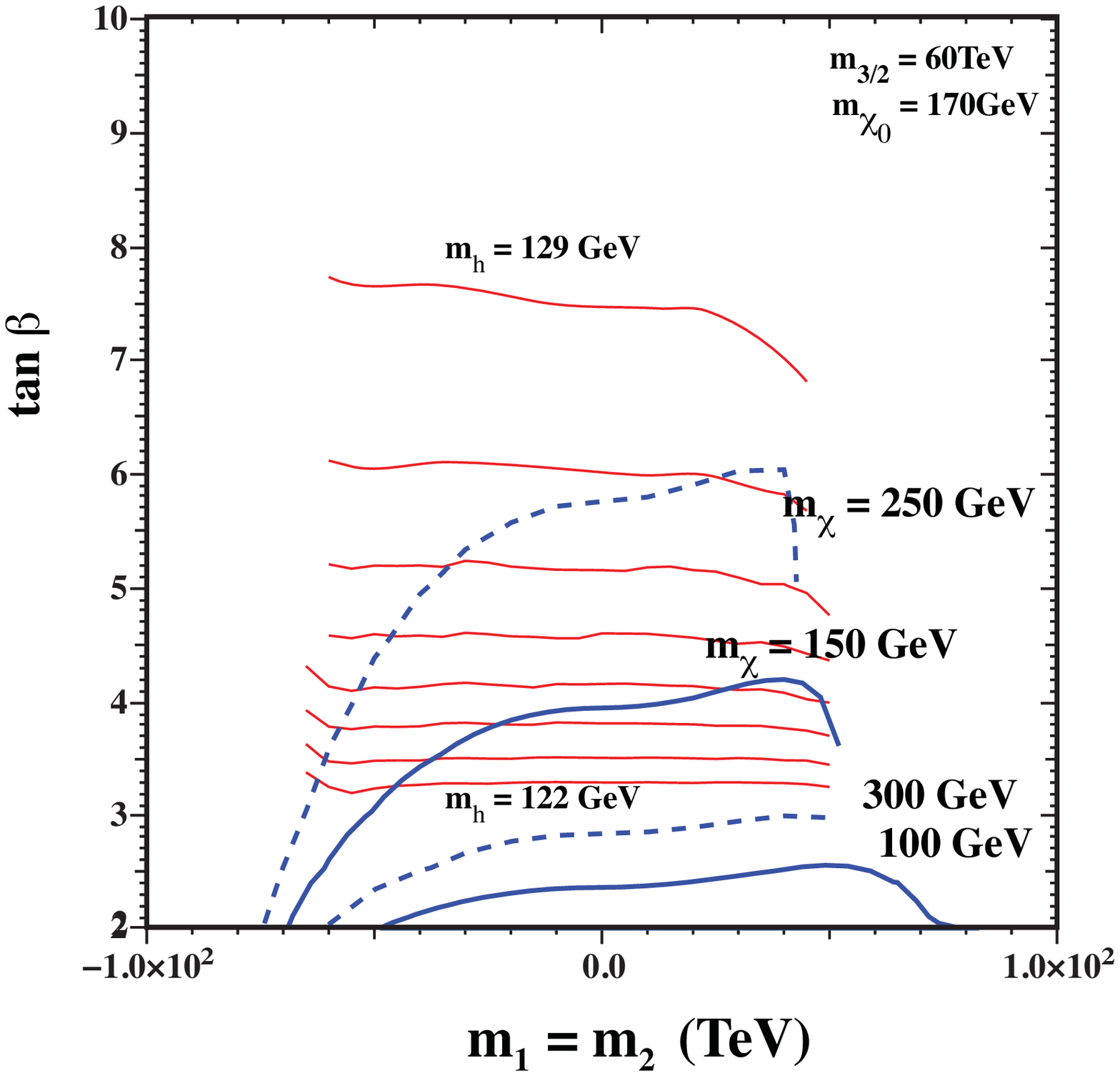,height=3.1in}
\epsfig{file=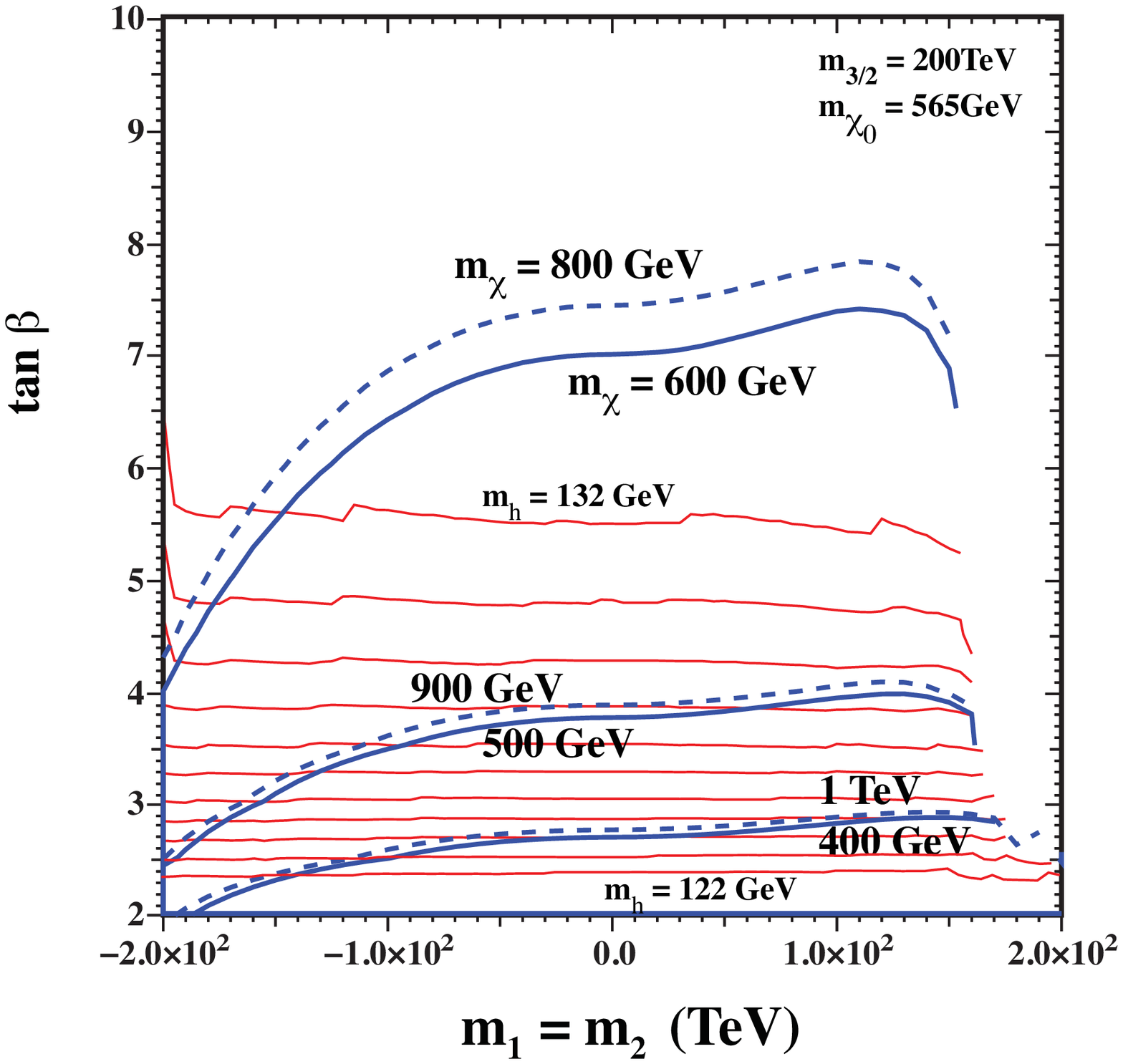,height=3.1in}
\hfill
\end{minipage}
\caption{
{\it
The $\tan \beta$--$m_{1,2}$ plane for a) $m_{3/2} = 60$\,TeV and b)   $m_{3/2} = 200$\,TeV.
The Higgs mass is shown by the nearly horizontal thin red contours in $1$\,GeV intervals.
The wino/chargino mass is shown by the thick solid ($\mu > 0$) and dashed
($\mu < 0$) contours.
}}
\label{fig:nuhm1}
\end{figure}

In Fig.\,\ref{fig:nuhm1}, we show two examples of ($\tan \beta$, $m_1=m_2$) planes
for fixed $m_{3/2}$.  In the left (right) panel, we fix $m_{3/2} = 60$ $(200)$\,TeV.
Shown are the contours of $m_h$ (red, nearly horizontal) and the wino mass, $m_{\chi}$,
for $\mu > 0$ (solid blue) and $\mu < 0$ (dashed blue).
The sign on the abscissa refers to the sign of $m_{1,2}^2$.
It should be noted that neither a gravitino mass of $60$\,TeV or $200$\,TeV
is consistent with $m_h \simeq 126\pm2$\,GeV in the universal case, i.e. $m_{1,2}=m_{3/2}$.
Higher gravitino masses are also possible with non-universal Higgs masses, but the results in that
case, do not differ significantly from results in the universal case.

For $m_{3/2}$ = 60 TeV, we see that the allowed range in $\tan \beta$ now extends from
$3.7$--$6.1$ due to the non-universality, which allows $m_h \simeq 126 \pm 2$ GeV for
this gravitino mass.
The range of $m_1 = m_2$ in the curves is limited by the electroweak symmetry breaking minimization conditions.
On the left, for large negative $m_{1,2}^2$, the Higgs pseudo-scalar mass-squared becomes negative
($m_A^2 < 0$).
On the right, at large positive $m_{1,2}^2$, $\mu^2 < 0$.
In either case, the model is inconsistent.
The figure shows that
$m_{3/2} = 60$\,TeV is not viable in universal PGM as emphasized above.
That is, $m_1 = m_2 = m_{3/2}$ is only possible for very low $\tan \beta < 2.7$,
where $m_h$ is too small.
As indicated on the plot, the contribution to the lightest neutralino from anomaly mediation is $170$\,GeV and is found using Eq. (\ref{eq:M2}).
At low $\tan \beta$, the threshold corrections from the Heavy Higgs bosons
and the Higgsinos increase the mass for $\mu < 0$ and decrease the mass for $\mu > 0$.
The effect is less pronounced at higher $\tan \beta$ and (although not shown in the left panel)
the radiative corrections are positive for both signs of $\mu$ at sufficiently large $\tan\beta$. The gluino mass in this case is $1.7$\,TeV, and would
be within the LHC reach.

The right panel of Fig.\,\ref{fig:nuhm1} shows the same plane when $m_{3/2} = 200$\,TeV
(a value still too low for universal PGM for $m_h \simeq 126\pm2$\,GeV).
In this case, the range in $\tan \beta$ is becoming compressed
(as in the universal model) and must lie in the range $2.7$--$3.6$ for $m_h \simeq 126 \pm 2$ GeV.
At large negative $m_{1,2}^2$, $m_A^2<0$, but this occurs for values beyond those displayed.
Again, we see that the universal case is realized at low $\tan \beta$ where the Higgs
mass contours extend to $m_1=m_2 = m_{3/2}$. For $m_{3/2}$ = 200\,TeV, the tree level
neutralino mass is $565$\,GeV, unfortunately the gluino mass is now $4.9$\,TeV, outside the
LHC reach. Here, we see that the threshold corrections for $m_\chi$ when $\mu > 0$ are positive when $\tan \beta$ is large.
For larger values of $m_{3/2}$, we are forced to a still tighter range in $\tan \beta$ until we eventually arrive at the universal case.

\section{NUHM2 models}
\begin{figure}[ht]
\begin{minipage}{8in}
\epsfig{file=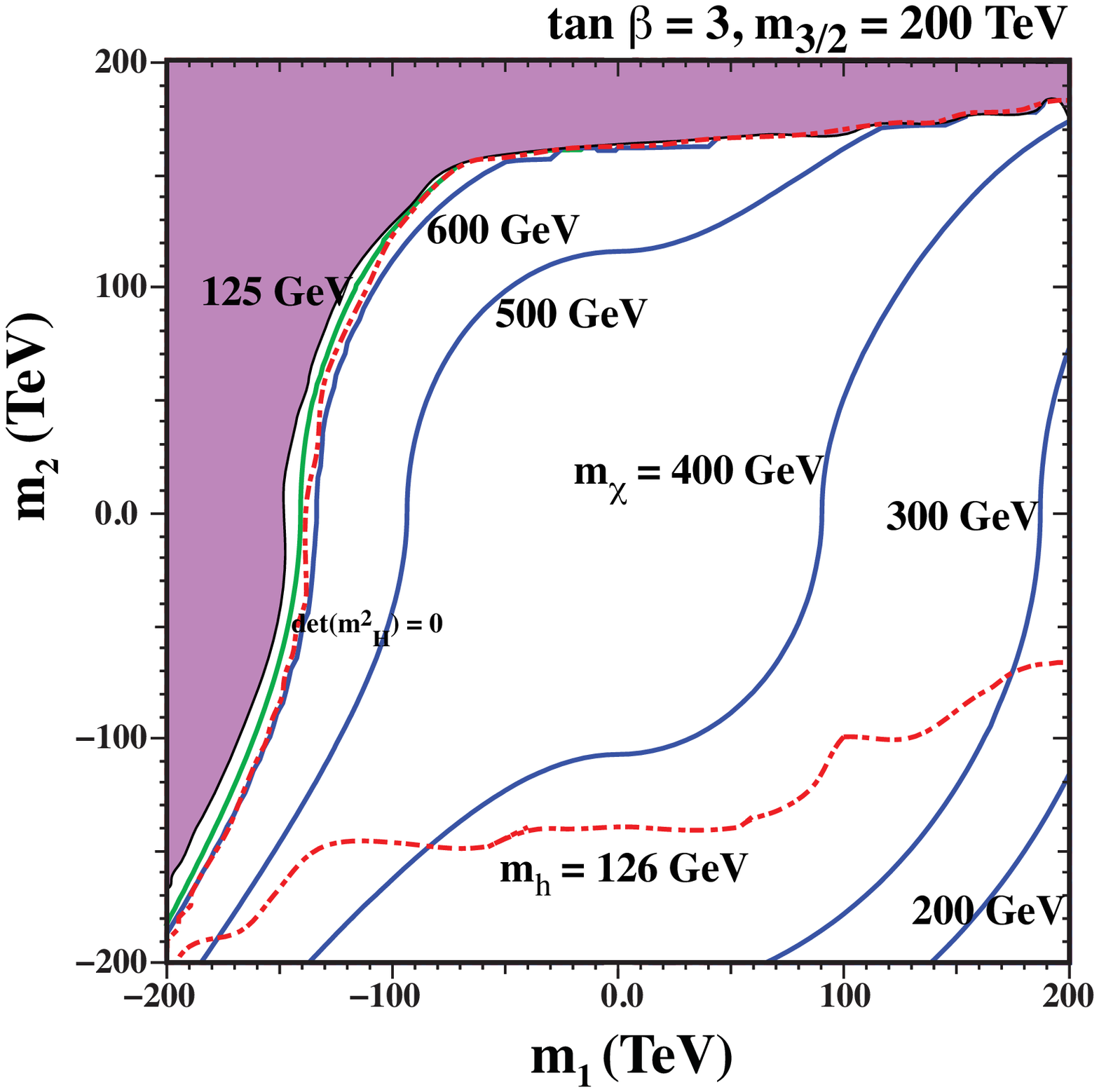,height=3.1in}
\epsfig{file=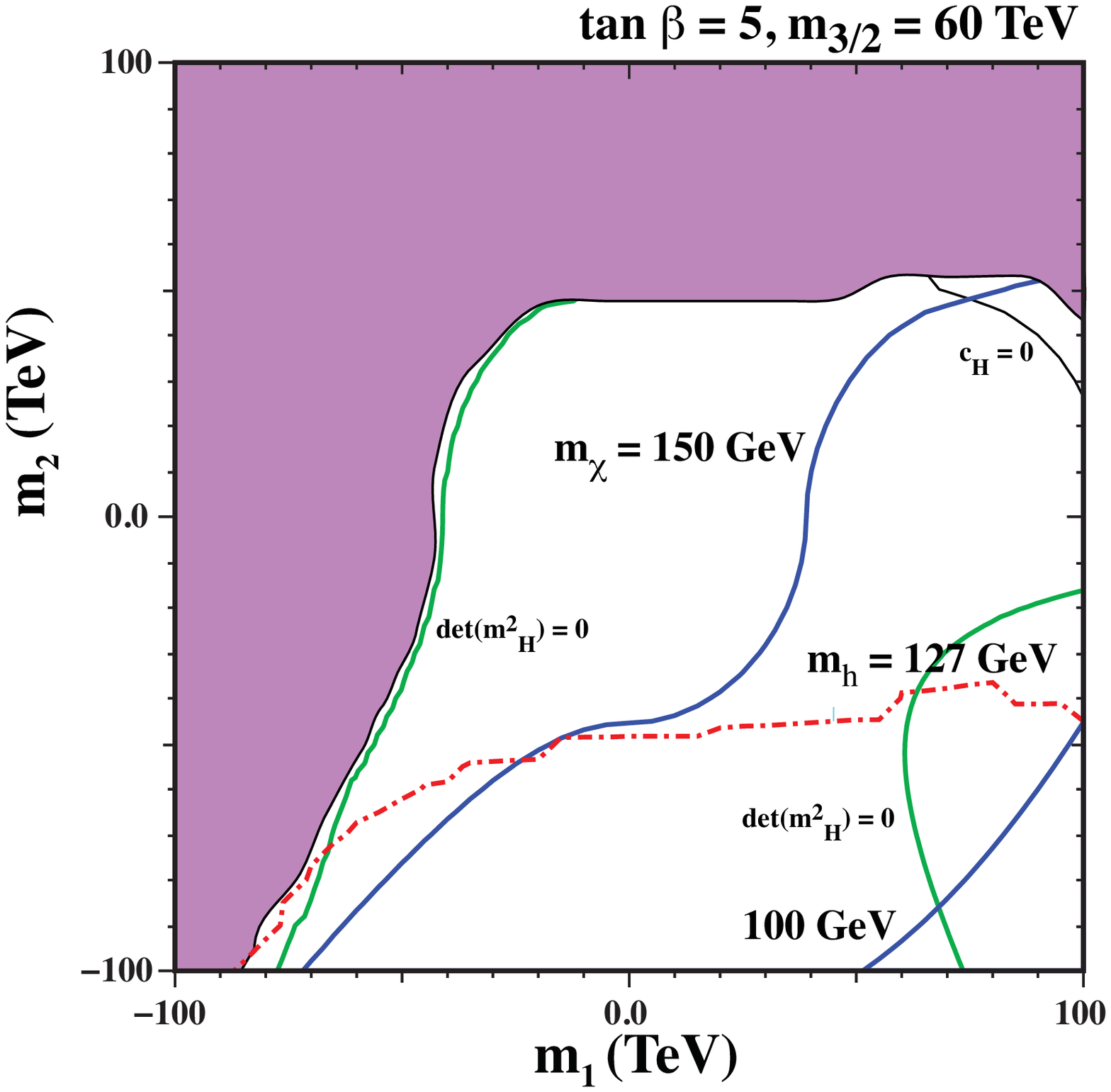,height=3.1in}
\hfill
\end{minipage}
\caption{
{\it
The $(m_1,m_2)$ plane for fixed $(m_{3/2},\tan \beta)$ = $(200$\,TeV, $3)$ (left) and $(60$\,TeV, $5)$ (right) for $\mu > 0$.
Shown are contours for the light Higgs mass, $m_h$, the lightest neutralino mass, $m_\chi$, and ${\det}(m_H^2)=0$.
The pink shaded region is excluded as either $\mu^2 < 0$ (horizontal region at large $m_2$)
or because $m_A^2< 0$ (vertical region at large negative $m_1$).
In the right panel, there is a specific combination
of $m_1$ and $m_2$ where $c_H = 0$ as seen by the black contour near the upper right corner of the figure.
}}
\label{nuhm2}
\end{figure}

One can further generalize the PGM model by allowing both Higgs soft masses
$m_1$ and $m_2$ to be independent of the gravitino mass.
Doing so, however, does not substantially expand the parameter space beyond the
NUHM1-type models of the previous section.
In Fig.\,\ref{nuhm2}, we show two examples of $(m_1, m_2)$ planes for fixed
$m_{3/2}$ and $\tan \beta$ with $\mu > 0$.
In the left panel, we show the plane for $m_{3/2} = 200$\,TeV and $\tan \beta = 3$.
Shown are contours of the lightest neutralino mass from $200$--$600$\,GeV.
Over much of the plane, the Higgs mass is close to $126$\,GeV.
The shaded region is excluded as radiative electroweak symmetry breaking conditions
can not be solved consistently there.
When $m_2^2$ is large and positive, $\mu^2 <0$,
and when $m_1^2$ is large and negative, $m_A^2 < 0$.
Near the shaded region, one sees a contour of $m_h = 125$\,GeV.

The green solid line corresponds to the ${\rm det}(m_H^2)=0$, where $m_H^2$
is the GUT scale Higgs squared mass matrix and is defined as
\begin{eqnarray}
m_H^2=\left(\begin{array}{cc} m_2^2+|\mu|^2 & -B\mu \\
-B\mu & m_1^2 +  |\mu|^2
\end{array}\right).
\end{eqnarray}
These contours correspond to regions where a linear combination of the Higgs bosons is a Nambu-Goldstone boson as was considered in \cite{Inoue:1985cw}.
This will be relevant for our later discussion about the Higgs bosons as a Nambu-Goldstone bosons.

In the right panel,  $m_{3/2} = 60$\,TeV and $\tan \beta = 5$.
As we have seen in the NUHM1-like models, larger $\tan \beta$ requires smaller $m_{3/2}$
and hence smaller neutralino masses.
In this case, $m_h$ is close to $127$\,GeV throughout the plane.
Again, the solid green line correspond to ${\det}(m_H^2)=0$.

For this choice of inputs, there is also the possibility of satisfying the
electroweak boundary conditions without the Higgs bi-linear term
in the K\"ahler potential.
The black solid line is where $c_H=0$ and hence the generalized GM term is not needed.
These points with $c_H=0$ are suggestive of a Peccei-Quinn symmetry.
In a theory symmetric under the PQ symmetry, both $c_H$ and $\mu$ are vanishing.
If the PQ symmetry is broken by the vev of some field $\langle \phi \rangle$ with vev of order $10^{12}$\,GeV, the axion could be the dark matter of the universe.
This field can also, through Planck suppressed operators
\begin{eqnarray}
W\supset \frac{\phi^2}{M_P}H_1H_2,
\end{eqnarray}
explain the origin of a $\mu$ parameter of order ${\cal O}(100)$\,TeV \cite{fy}.

\begin{figure}[h!]
\begin{minipage}{8in}
\epsfig{file=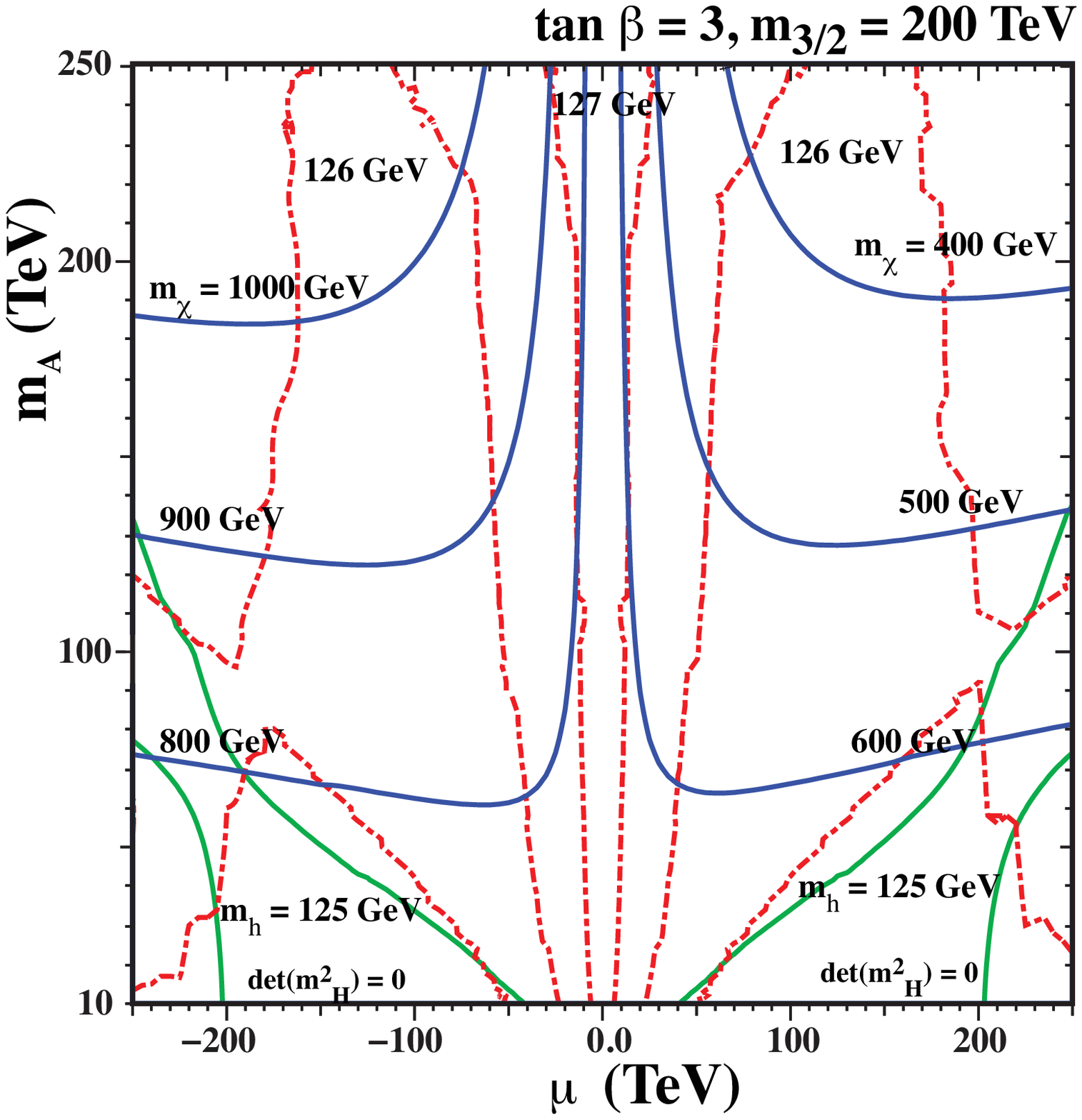,height=3.1in}
\epsfig{file=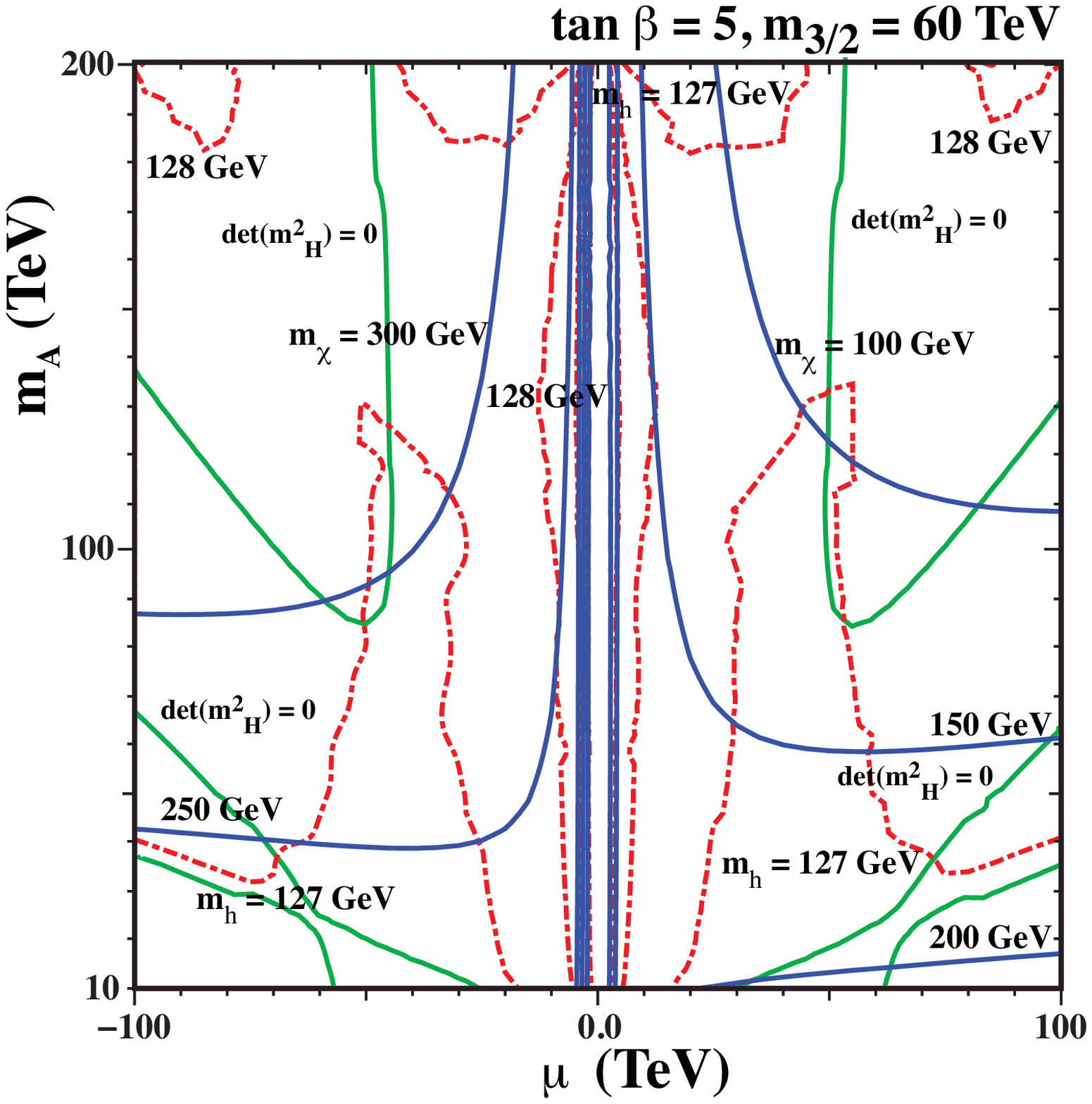,height=3.1in}
\hfill
\end{minipage}
\begin{minipage}{8in}
\epsfig{file=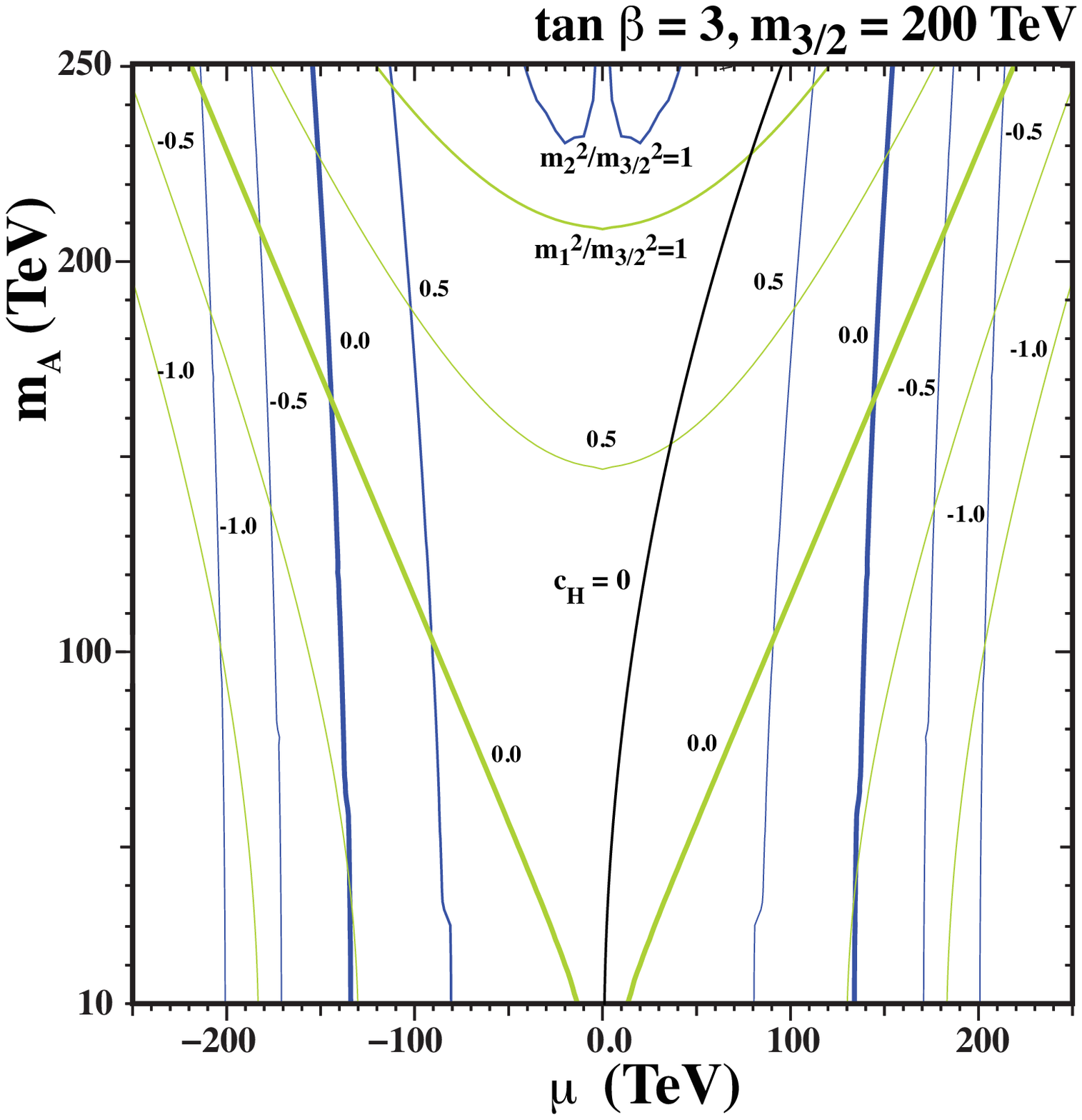,height=3.1in}
\epsfig{file=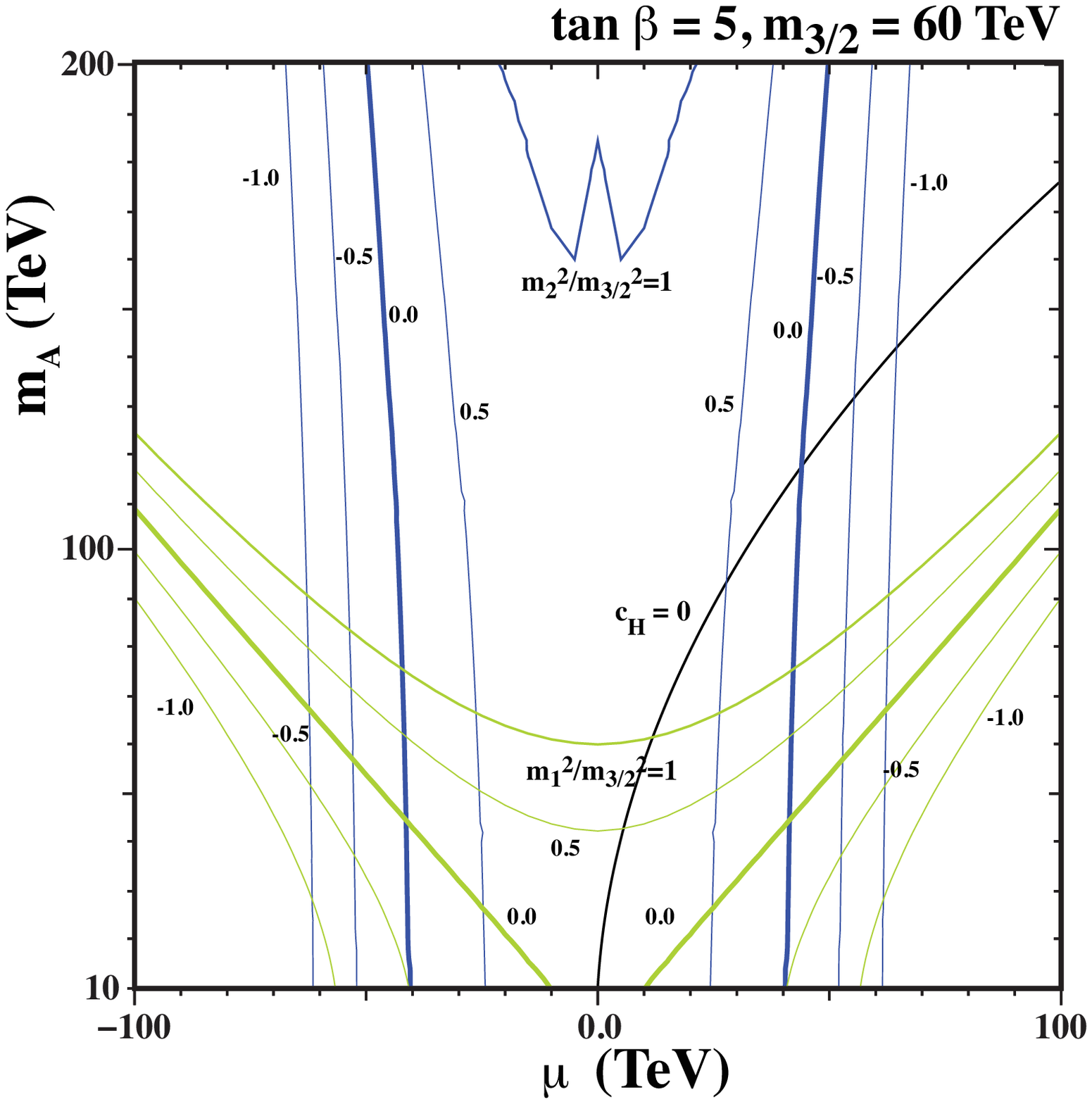,height=3.1in}
\hfill
\end{minipage}

\caption{
{\it
The $(\mu, m_A)$ plane for fixed $(m_{3/2},\tan \beta)$ = $(200$\,TeV$, 3)$ (left)
and $(60$\,TeV$, 5)$ (right).
Shown in the upper panels are contours for the light Higgs mass, $m_h$ (red, dot dashed)and the lightest neutralino mass, $m_\chi$ (solid blue). The contour where the determinant of the Higgs squared  mass matrix is  $0$ at the GUT scale is shown by the solid green curve as labeled.
Shown in the lower panels are contours for $m_1^2/m_{3/2}^2$ (orange) and  $m_2^2/m_{3/2}^2$ (blue)
for $m_i^2/m_{3/2}^2 = 1$ (medium), $0$ (thick) and $0.5$, $-0.5$ and $-1$ (thin).
The contour where the Higgs coupling $c_H = 0$ is shown by the solid black curves as labeled.
}}
\label{mumA}
\end{figure}

An alternative way of viewing the NUHM2 parameter space is through a ($\mu, m_A$) plane.
One can equally well choose the parameters of the NUHM PGM model as
$m_{3/2}, \tan \beta, \mu$, and $m_A$, the latter two replacing the Higgs soft masses
which are now determined at the weak scale by the electroweak boundary conditions and then
run back up to the GUT input scale.  We show the same two examples from Fig.\,\ref{nuhm2}, but now in the ($\mu, m_A$) plane in
Fig.\,\ref{mumA}. As in Fig.\,\ref{nuhm2}, we show contours of the lightest neutralino mass (solid blue)
and of the Higgs mass (red dot-dashed) in the upper two panels. The Higgs mass again changes relatively slowly across these planes due its relative insensitivity to $\mu$ and $m_A$ once $m_{3/2}$ and $\tan \beta$ are fixed.
The green lines show where ${\det} (m_H^2)=0$.
In the lower two panels, we also show the contours of $m_1^2/m_{3/2}^2$ (orange)
and  $m_2^2/m_{3/2}^2$ (blue) for $m_i^2/m_{3/2}^2$ = $1$ (medium), $0$ (thick) and $0.5$, $-0.5$
and $-1$ (thin).
A universal
PGM model would require that the contours for $m_1^2/m_{3/2}^2 = m_2^2/m_{3/2}^2 = 1$
intersect at some point.  Comparing the two panels, we see that an intersection is almost found
for $\tan \beta = 3$.  In fact, from our previous study of the universal PGM model, we
expect that such an intersection would occur around $\tan \beta = 2.5$ for $m_{3/2} = 200$\,TeV.
The black line shows the relation between $\mu$ and $m_A$ where $c_H=0$.

\section{Higgs doublets as Nambu-Goldstone Bosons}
It is very interesting that results in the previous section show
that the SUSY breaking soft masses for $H_2$ and/or $H_1$ can vanish at the GUT scale. As was discussed in Section \ref{PGMnU}, this could be due to a no-scale like K\"ahler manifold for the Higgs bosons. However, another intriguing possibility is that one or both of the Higgs multiplets are
pseudo Nambu-Goldstone (NG) fields arising from some unknown physics at the GUT scale.%
\footnote{
The idea of the Higgs as a pseudo-Goldstone boson was established long
ago\,\cite{Kugo:1983ai}.
The minimal non-linear sigma model with one of the Higgs doublets as a pseudo
Nambu-Goldstone multiplet was discussed in Ref.\,\cite{GY},
where the non-linear sigma model couples to supergravity without spoiling the symmetry.
Non-linear sigma models where not only the Higgs doublets
but also other matter multiplets are pseudo Nambu-Goldstone multiplets were discussed in Ref.\,\cite{Buchmuller:1983na}.
}
 If this is indeed the case, the NG hypothesis provides a natural answer to two fundamental questions. One is their small Yukawa couplings to quarks and leptons at the GUT scale
and the other is their vanishing $R$ charges.
The small Yukawa couplings are easily understood because of the celebrated low energy theorem
for NG fields.
The vanishing $R$ charges is also inevitable for the NG fields
since they transform non-linearly under the broken symmetry.

It is known that the coset space for the NG fields must be a K\"ahler manifold in SUSY theories.
Thus, let us discuss possible K\"ahler manifolds which accommodate one or two Higgs multiplets as NG fields.
First, we consider the K\"ahler manifold, $SU(4)/SU(2)\times SU(2)\times U(1)$ which is the minimal coset space for two Higgs doublets, $H_2$ and $H_1$.
The coset space, $SU(4)/SU(2)\times SU(2)\times U(1)$, has a broken generator, $X(2,2)_{+1}$ and its complex conjugate and hence we have corresponding NG chiral multiplets $\Phi$
which transform as the broken generator $X(2,2)_{+1}$ under the unbroken symmetry.
 Now we identify the first $SU(2)$ with the electroweak gauge symmetry $SU(2)_L$ and the $U(1)$ subgroup in the second $SU(2)$ with the hypercharge gauge symmetry $U(1)_Y$ in the standard model. Then, it is clear that the NG multiplets $\Phi$
 are nothing but the Higgs doublets, $H_2$ and $H_1$. The last $U(1)$ symmetry in the unbroken group must be a global symmetry since it is anomalous. Notice that both of the Higgs multiplets carry +1 charge of this global symmetry and hence it may be identified with the Peccei-Quinn $U(1)$.

Now we couple the above non-linear sigma model to supergravity. It was pointed out that any compact
K\"ahler manifold cannot be coupled to supergravity \cite{KS, KY}. There have been two solutions proposed so far to solve this problem. One is to introduce a new massless modulus field \cite{KS} and the other is to break the PQ $U(1)$ subgroup in the unbroken, $SU(2)\times SU(2)\times U(1)$ \cite{KY}.%
\footnote{
In no-scale supergravity\,\cite{Cremmer:1983bf},
the SUSY breaking and matter fields are the coset space variables
of a non-compact K\"ahler manifold $SU(n,1)/SU(n)\times U(1)$.
The $SU(n,1)$ symmetry is, however, explicitly broken in supergravity. In spite of this, the
soft masses of the matter fields are still vanishing due to the sequestered structure
of the K\"ahler potential.
In this paper, we do not pursue no-scale models since they predict vanishing
anomaly mediated SUSY breaking effects (see e.g. \cite{Izawa:2010ym}).
Nevertheless, sequestering only the Higgs fields as in Eq.\,(\ref{K2}) and in the appendix
can account for the vanishing of the Higgs soft masses alone.
}
In the former case, we need some extra mechanism to break the PQ $U(1)$ subgroup.
In the latter case, the coset space is $SU(4)/SU(2)\times SU(2)$ and we have one additional NG multiplet $\phi$. The imaginary part of the complex boson in $\phi$ is nothing but the axion and the real part is identified with the
so-called saxion. The decay constant $F_a$ may be at the GUT or Planck scale.

The K\"ahler potential for the $SU(4)/SU(2)\times SU(2)$ non-linear sigma model is written as \cite{KUY}
\begin{eqnarray}
 K = - M_{\rm *}^2 F( \det \xi^{\dagger i}_{a} \xi^a_j ) \ ,
\end{eqnarray}
where $F$ is an arbitrary function, $M_* \simeq M_{\rm GUT}$ or $M_{\rm PL}$,
while $\xi$ is the dimensionless variable parametrizing the coset space
related to $\Phi$ and $\phi$ via
 \begin{eqnarray}
 \label{eq:rep}
 \xi^a_i  =
 \left(
\begin{array}{c}
e^{\phi}\mathsf{1}_{2\times2}\\
 \Phi
\end{array}
\right)\ .
\end{eqnarray}
The index $i$ corresponds to the first unbroken $SU(2)$, $i = 1$--$2$ while $a$  runs from $1$--$4$.
As we mentioned above,  $\phi$ and $\Phi$ transform under the $U(1)(\subset SU(4))$ PQ
symmetry,
\begin{eqnarray}
 \phi \to \phi' =  \phi  - i \alpha\ , \quad \Phi \to \Phi' = e^{i\alpha} \Phi\ ,
\end{eqnarray}
and hence, $\phi$ plays the role of the  axion multiplet with the decay constant $F_a = {\cal O}(M_{\rm GUT})$.

It should be noted that  $\det \xi^{\dagger i}_{a} \xi^a_j$ is invariant
under the non-linear $SU(4)$ transformation,
\begin{eqnarray}
 \xi \to \xi' = g\, \xi h^{-1}(g, \xi)\ ,
\end{eqnarray}
where $g\in SU(4)$ and $h(g,\xi)$ is a local $SU(2)$ matrix superfield \cite{KY}.
Thus, since K\"ahler potential itself is invariant under the $SU(4)$ symmetry, the non-linear sigma-model
can be safely coupled to the supergravity without spoiling the symmetry\,\cite{KY}.
In particular, all the scalars in the Higgs doublets in $\Phi$ are Goldstone bosons of
$SU(4)/SU(2)\times SU(2)\times U(1)$, and hence,
they remain massless even after supersymmetry breaking.
The axion multiplet $\phi$, on the other hand, contains the so-called the quasi-Goldstone boson (saxion)
which is not truly associated with the PQ-symmetry.
As a result, it  obtains a soft mass $m_{\rm saxion}^2=m_{3/2}^2$ (see \cite{GY} for details).

By remembering that the K\"ahler manifold $SU(4)/SU(2)\times SU(2)\times U(1)$
is equivalent to $U(4)/U(2)\times U(2)$, one may also take the coset variables of  $U(4)/SU(2)\times SU(2)\times U(1)$ instead of those of $SU(4)/SU(2)\times SU(2)$,
which can also be coupled to supergravity without spoiling the symmetry.
There, $\phi$ corresponds to a Goldstone mode of the non-linearly realized
$U(1)$ symmetry, while the remaining $U(1)$ symmetry (under which $\Phi$ rotates)
is the PQ-symmetry which is assumed to be broken spontaneously at low energies
at around $10^{12}$\.GeV and leads to the axion.

As an interesting possibility for the model of  $U(4)/SU(2)\times SU(2)\times U(1)$,
the true goldstone boson in $\phi$ can play the role of quintessence
and explain the observed dark energy\,\cite{Nomura:2000yk}. This occurs when the non-linearly realized
$U(1)$ shift of $\phi$ is nearly exact, being  broken only by its anomalous couplings to the gauge kinetic terms as for the string axion.
In this case, a linear combination of the axion and the true Goldstone boson of $\phi$
are free from the QCD anomaly and obtains a mass only from the $SU(2)_L$ instantons.
Therefore, the mass of the corresponding mode is extremely small and can play
the role of quintessence in explaining dark energy.

Once we realize the Higgs doublets (and the axion) supermultiplets as the NG modes of
$SU(4)/SU(2)\times SU(2)$ (or $U(4)/SU(2)\times SU(2)\times U(1)_{PQ}$),
 we introduce the MSSM gauge interactions,  the Yukawa interactions,
and the bi-linear term of the Higgs doublets discussed in the previous section
as perturbative explicit breaking to the $SU(4)$ symmetry.
Therefore, the NG hypothesis of the Higgs doublets consistently works with the pure gravity mediation model.

So far, we have considered that both of the Higgs doublets appear as NG fields in the above realization
so that $m_{1}^2 = m_{2}^2 = 0$.
One may also take a relaxed NG Higgs hypothesis where only one of the Higgs doublets appear
as a NG multiplet by considering a smaller coset space.
For example, a smaller coset space $U(3)/SU(2)\times U(1)$
accommodates only one of the Higgs doublets as a NG field.
Here, the unbroken group is identified with the electroweak
gauge group, $SU(2)_L\times U(1)_Y$.
We also have an axion-like NG multiplet $\phi$ as in the case of $U(4)/SU(2)\times SU(2)\times U(1)$.
In this case, either $H_2$ or $H_1$ can be identified as the NG mode and has vanishing soft masses
while the other Higgs doublet obtains a mass of $O(m_{3/2}^2)$ once we couple it to supergravity.

As another realization of this less constraining NG Higgs hypothesis,
it is also possible to identify the two Higgs doublets as the coset space variables of
 $U(3)/SU(2)\times U(1)$ in the so-called doubling realization \cite{bando,kugo}.
In the doubling realization of  $U(3)/SU(2)\times U(1)$,
we introduce a pair of representative coset space variables
\begin{eqnarray}
\xi =
\left(
\begin{array}{ccc}
1\\
H_1
\end{array}
\right)\ , \quad
\bar{\xi} = (1, H_2)\ ,
\end{eqnarray}
instead of the one in Eq.\,(\ref{eq:rep}). Here, $H_1$ and $H_2$ are implicitly normalized by $M_*$ so that they are dimensionless.
Under the non-linearly realized $U(3)$ symmetry, they transform as
\begin{eqnarray}
\xi\to \xi' = g \xi v^{-1}\ , \quad
\bar\xi\to \bar\xi' = \bar{v}^{-1}\xi g^\dagger\ ,
\end{eqnarray}
where $g\in U(3)$ and $v\, (\bar{v})$ are chiral superfields representing a local $U(1)$ transformation.
By using these, we can construct a $U(3)$ invariant K\"ahler potential \cite{Goto:1992hh},
\begin{eqnarray}
K = M_*^2F(x)\ ,
\end{eqnarray}
where $x$ is the $U(3)$ invariant combination,
\begin{eqnarray}
x = \xi^\dagger \xi (\bar\xi\xi)^{-1}\bar\xi^\dagger \bar\xi (\bar\xi^\dagger\xi^\dagger)^{-1}
  = \frac{(1+H_1^\dagger H_1)(1+H_2^\dagger H_2)}
   {(1+H_1 H_2)(1+H_1^\dagger H_2^\dagger)}\ .
\end{eqnarray}
Thus, we can couple the non-linear sigma model of $U(3)/SU(2)\times U(1)$ to supergravity
without spoiling the symmetry.

In this example, although both Higgs doublets appear as NG multiplets,
only one linear combination is a true NG mode while the other
is a quasi NG mode.
Thus, again, we have only one massless Higgs doublet after supersymmetry breaking.
It should be noted, however, that in this realization, the Higgs bi-linear term appears
in the K\"ahler potential without spoiling the symmetry.
Therefore, the vanishing scalar mass of the true NG mode means neither $m_{1}^2  = 0$
nor $m_2^2 = 0$. Instead, the model predicts $\det m_H^2 = 0$.

In what follows we examine some of the phenomenological aspects of the NG
hypothesis with one or both of the Higgs as NG states.

\subsection{One Higgs NG state}
\begin{figure}[ht]
\begin{minipage}{8in}
\epsfig{file=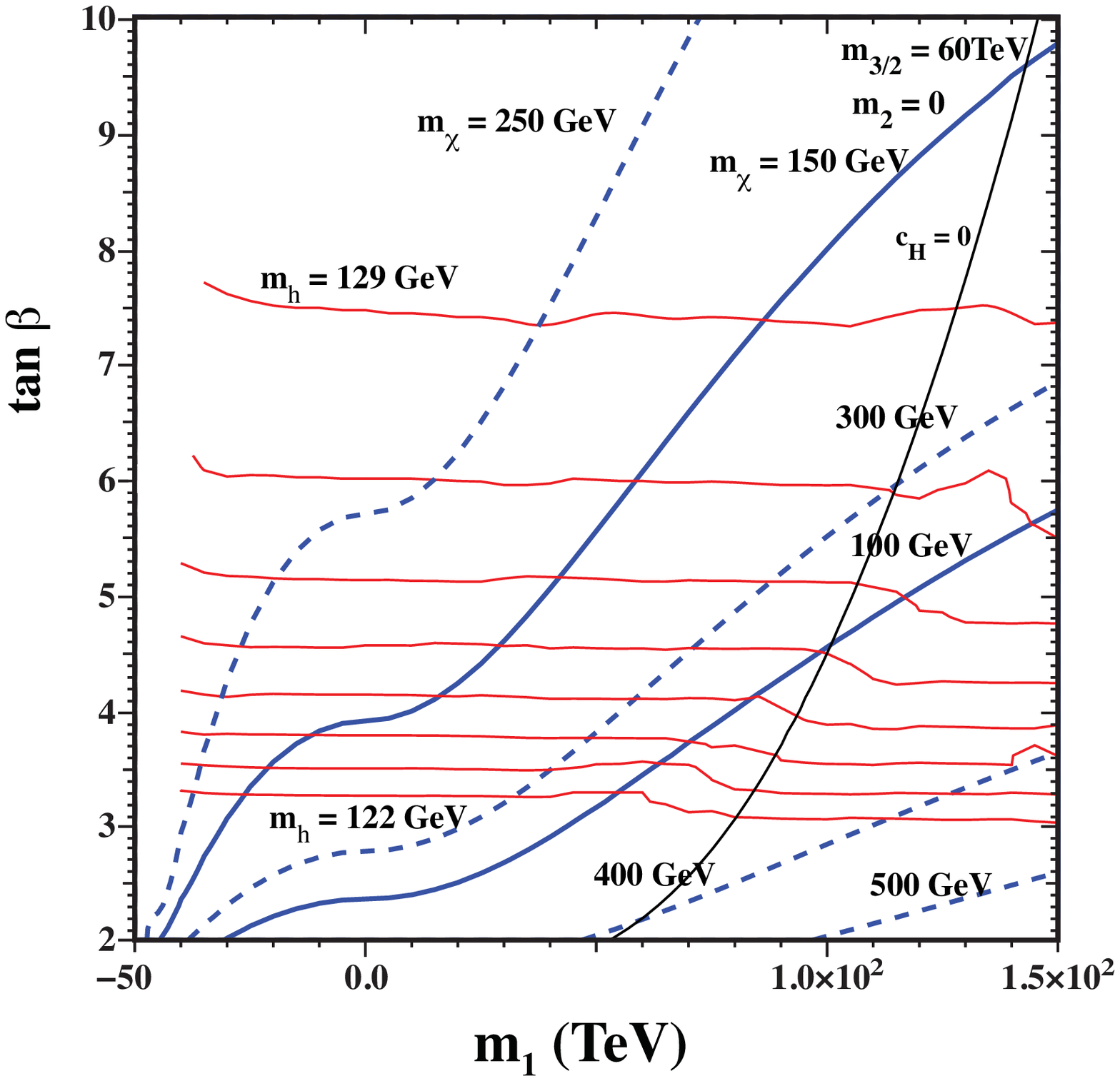,height=3.1in}
\epsfig{file=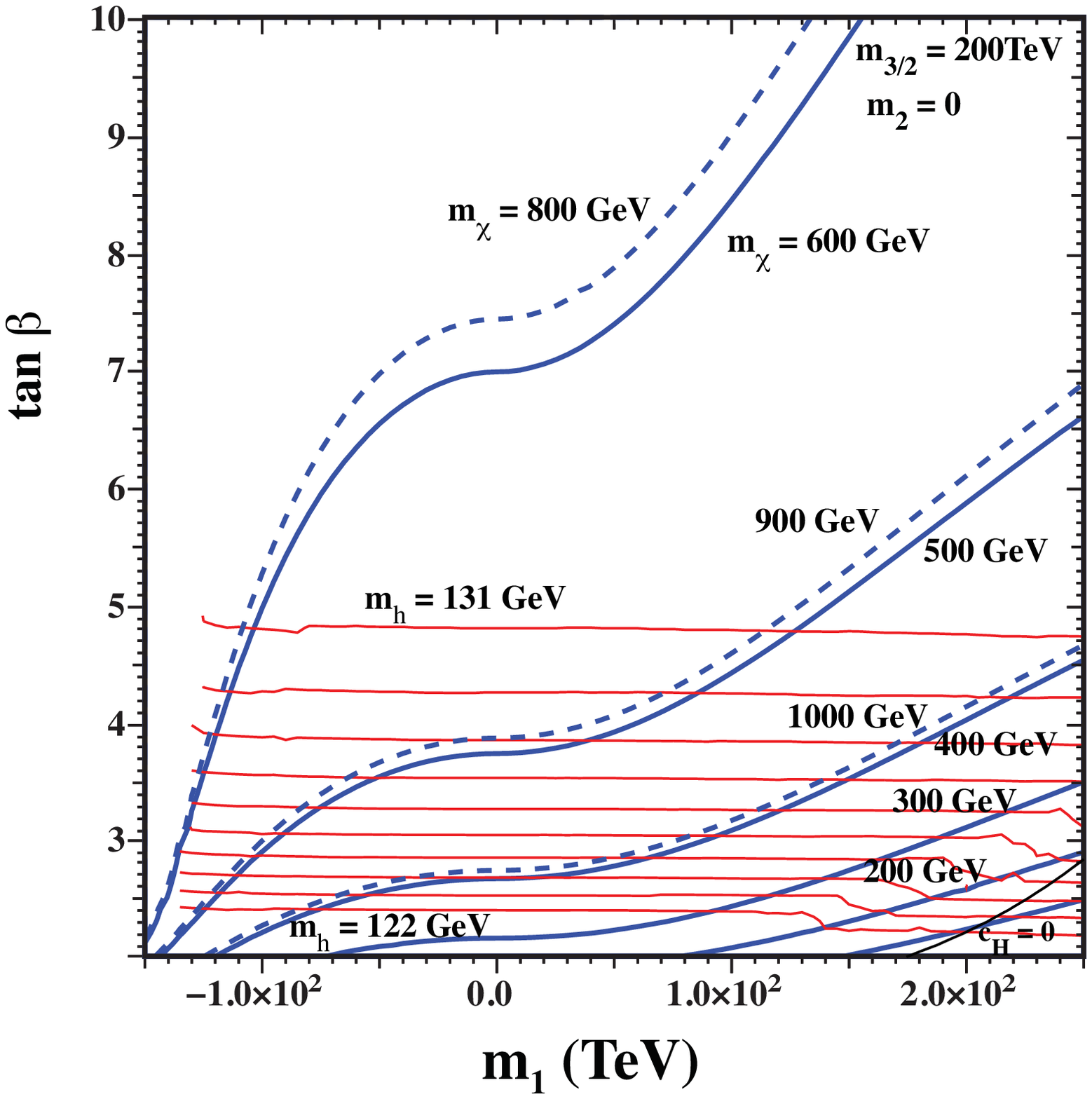,height=3.1in}
\hfill
\end{minipage}
\caption{
{\it
The $(m_1,\tan \beta)$ plane with $m_2 = 0$ for fixed $m_{3/2} = 60$\,TeV (left) and $200$\,TeV (right).
Shown are contours for the light Higgs mass, $m_h$ and the lightest neutralino mass, $m_\chi$.
The contour for $c_H = 0$ is shown by the solid black curve.
}}
\label{nuhm2m1}
\end{figure}

\begin{figure}[h!]
\begin{minipage}{8in}
\epsfig{file=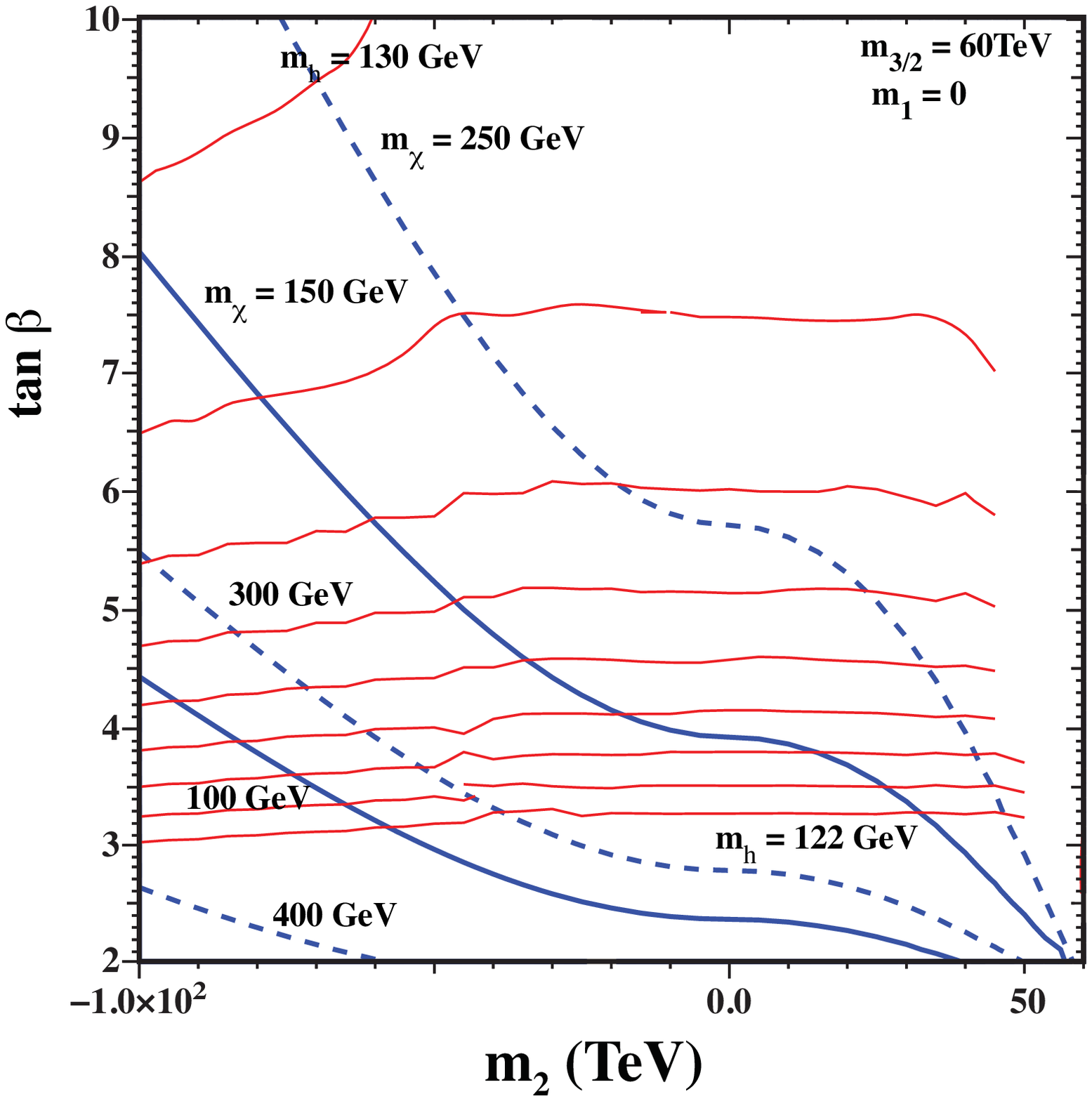,height=3.1in}
\epsfig{file=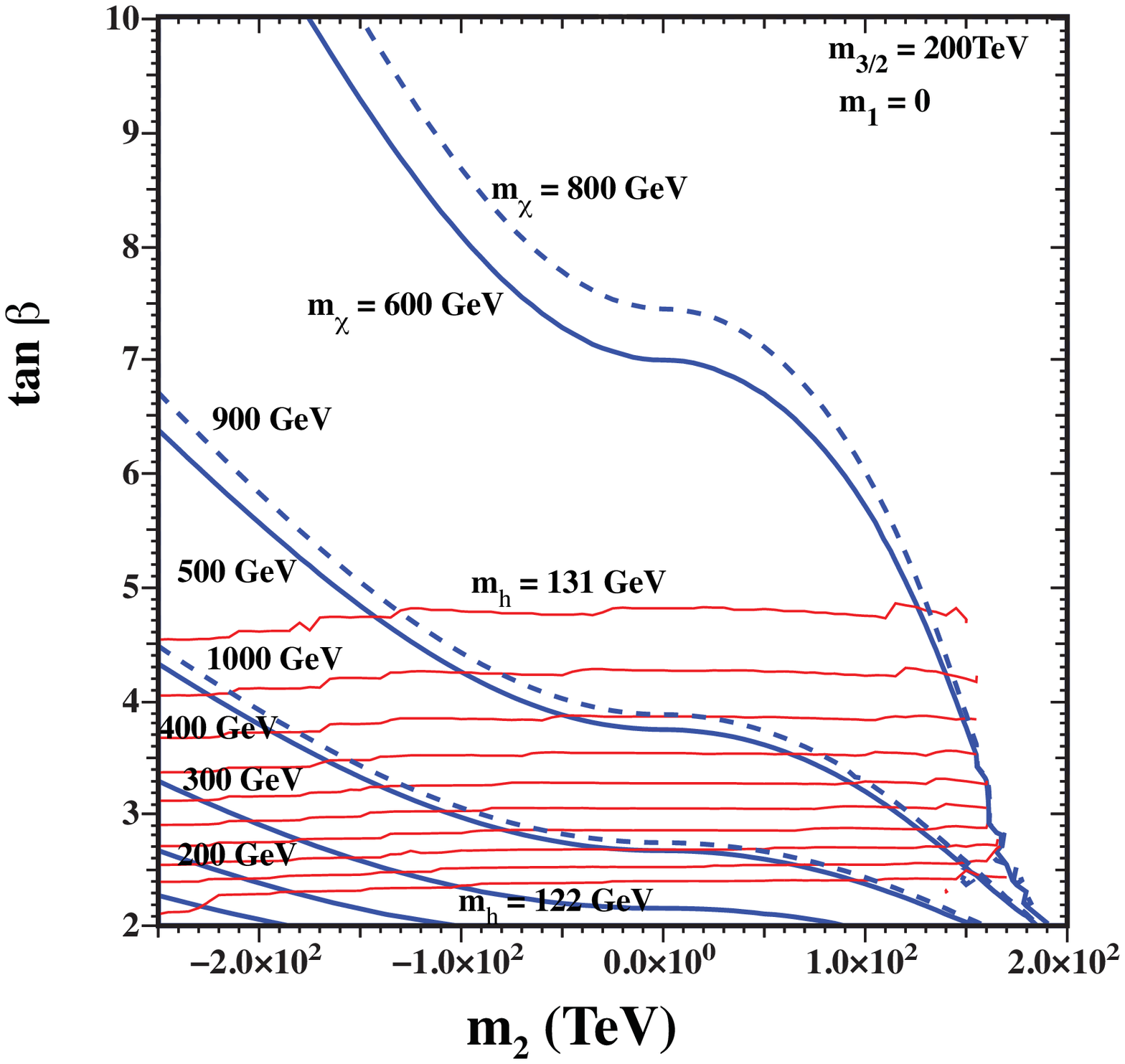,height=3.1in}
\hfill
\end{minipage}
\caption{
{\it
The $(m_2,\tan \beta)$ plane with $m_1 = 0$ for fixed $m_{3/2} = 60$\,TeV (left) and $200$\,TeV (right).
Shown are contours for the light Higgs mass, $m_h$ and the lightest neutralino mass, $m_\chi$.
}}
\label{nuhm2m2}
\end{figure}

In this section, we show our results for the case where one of the Higgs bosons is a pseudo Nambu-Goldstone boson.
As before, $\tan\beta$ is free to vary as long as $m_2$ is slightly smaller than $m_{3/2}$. Our results, when either $m_2 = 0$ or $m_1 = 0$, are shown in Fig.\,\ref{nuhm2m1} and \ref{nuhm2m2} respectively.
These figures also show the behavior of the lightest Higgs boson mass which is again predominantly determined by $m_{3/2}$ and $\tan\beta$.
As one can see, for the relatively low value of
$m_{3/2} = 60$\,TeV, $\tan \beta$ must lie in the range $3.5$--$6.2$ to obtain $m_h \simeq 126 \pm 2$\,GeV.
At the higher value of $m_{3/2} = 200$\,TeV, as we have seen before, $\tan \beta$ is constrained
to lower values, $2.4$--$3.6$.
Contours terminate on left, in each panel, at large negative $m_1^2$ when
$m_A^2 < 0$.

Fig.~\ref{nuhm2m1} also shows a contour in both panels where $c_H=0$ (black). For this line,
we do not need the Higgs bilinear term
in the K\"ahler potential. As was previously discussed this can be motivated by dark matter and the generation of a $\mu$ term of order $100$ TeV. However, for the NG hypothesis there is an additional motivation.
To get a phenomenologically viable model for the NG hypothesis, we needed to introduce explicit breaking of the $SU(4)$ symmetry in the form of: gauge interactions, the Yukawa couplings for the MSSM fields, and a bilinear term for the Higgs.  In the parameter space for which $c_H\ne 0$, we have also added an additional explicit breaking in the form of the generalized GM term.
For the contours with $c_H=0$, the tree-level K\"ahler potential is invariant under the $SU(4)$ symmetry. This contour is not seen in Fig.\,\ref{nuhm2m2} for the range of parameters considered.

The behavior of the lightest supersymmetric particle (LSP) mass, is more involved.
In this case, the wino is the LSP.
The dominant threshold correction to the LSP mass, comes from integrating out Higgsinos, which gives
\begin{eqnarray}
\Delta M_2 \simeq -\mu\frac{g^2}{16\pi^2}\sin2\beta\frac{m_A^2}{m_A^2-\mu^2}\ln\frac{m_A^2}{\mu^2}.\label{M2Thresh}
\end{eqnarray}
The $\tan\beta$ dependence can be seen in both Fig.\,\ref{nuhm2m1} and Fig.\,\ref{nuhm2m2}.
As $\tan\beta$ becomes larger, the deviation in $m_{\chi}$ from its anomaly mediated value,
$170$\,GeV for $m_{3/2}=60$\,TeV and $565$\,GeV for $m_{3/2}=200$\,TeV, is diminished.
To understand the $\mu$ dependence of the LSP mass, we look at the minimization condition for the Higgs potential.  As long as $\tan\beta$ is not too small, we have roughly $\mu^2\simeq -m_2^2$ at the weak scale.  Because the GUT scale value of $m_2=0$ is not varied, $\mu$ will be roughly constant in Fig.\,\ref{nuhm2m1} and so changes to $\mu$ will only have a minor affect on the mass of the LSP.
The pseudo scalar mass, $m_A$, on the other hand,  depends on $m_1$,
\begin{eqnarray}
m_A^2 \simeq m_1^2+m_2^2+2\mu^2,
\end{eqnarray}
and so will vary across the parameter space found in Fig.\,\ref{nuhm2m1}.
This variation in $m_A$ accounts for the dependence of the LSP mass on $m_1$.

In Fig.\,\ref{nuhm2m2}, we have taken $m_1=0$ and varied $m_2$.
The $\tan\beta$ dependence can be seen by comparing the wino masses with the anomaly mediated contribution.
It again roughly follows from Eq.\,(\ref{M2Thresh}).
The $m_A$ and $\mu$ dependence is now more complicated since they both vary.
However, as would be expected by examining Eq.\,(\ref{M2Thresh}), the $\mu$ dependence wins out.
Since $\mu$ becomes larger as $|m_2|$ becomes large, the deviation of the wino masses from the anomaly mediation value increases as $m_2$ becomes more negative.

\begin{figure}[h!]
\begin{center}
\epsfig{file=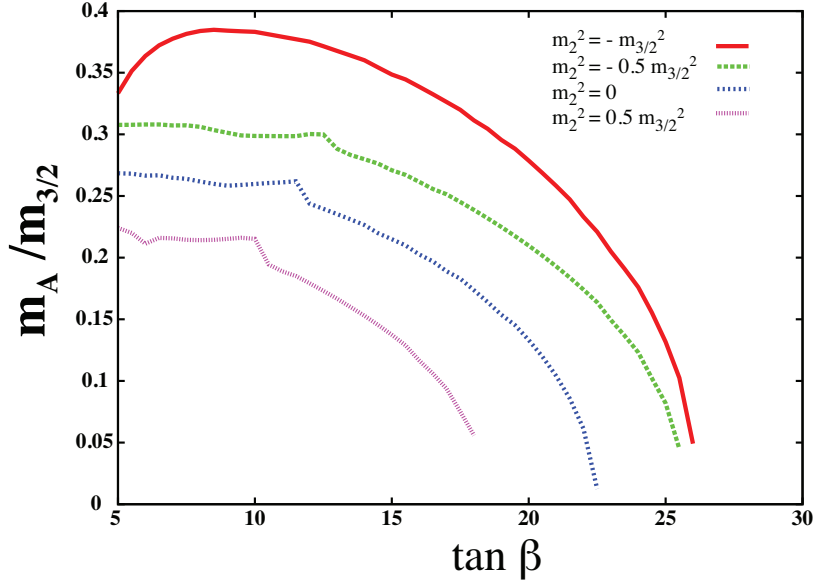,height=3.1in}
\end{center}
\caption{\it The mass of the pseudo-scalar as a function of $\tan\beta$ for ${\rm det}(m_H^2)=0$ at the
GUT scale with the given values of $m_2^2(M_{GUT})$ and $m_{3/2}=30$\,TeV.}\label{fig:DetMmA}
\end{figure}

As was depicted in Fig.\,\ref{nuhm2}, there are regions of parameter space for which some linear combination of the Higgs bosons is a pseudo Nambu-Goldstone boson at the GUT scale, i.e. ${\rm det}(m^2_H)=0$,
without any explicit breaking from the $\mu_0$ term.
The most interesting case occurs for large negative values of $m_1$. In this region, $m_A$ is quite small.  In fact, for specific values of $\tan\beta$ and $m_1^2$, $m_A$ can be within the reach of the LHC as can be seen in Fig.\,\ref{fig:DetMmA}. As we see in this figure, for larger negative values of $m_2$, $m_A$ is larger.  This can be attributed to the fact that $\mu^2\simeq -m_2^2$. Since $m_A$ is more sensitive to $\mu$ than $m_2^2$, $m_A$ becomes larger as $m_2^2$ becomes more negative.  The $\tan\beta$ effect can be traced back to the $m_2^2$ mass as well.  As $\tan\beta$ becomes larger, the top Yukawa coupling, $y_t$ becomes smaller.  For smaller values of $y_t$ $m_2$ is not driven as negative and so $m_A$ decreases.

\subsection{Two Higgs NG states}
In the case of both Higgs doublets arising as Nambu-Goldstone boson states,
we are reduced once more to an effective two-parameter theory
with $m_{3/2}$ and $\tan \beta$ as free parameters as in the universal model.

\begin{figure}[ht]
\begin{minipage}{8in}
\epsfig{file=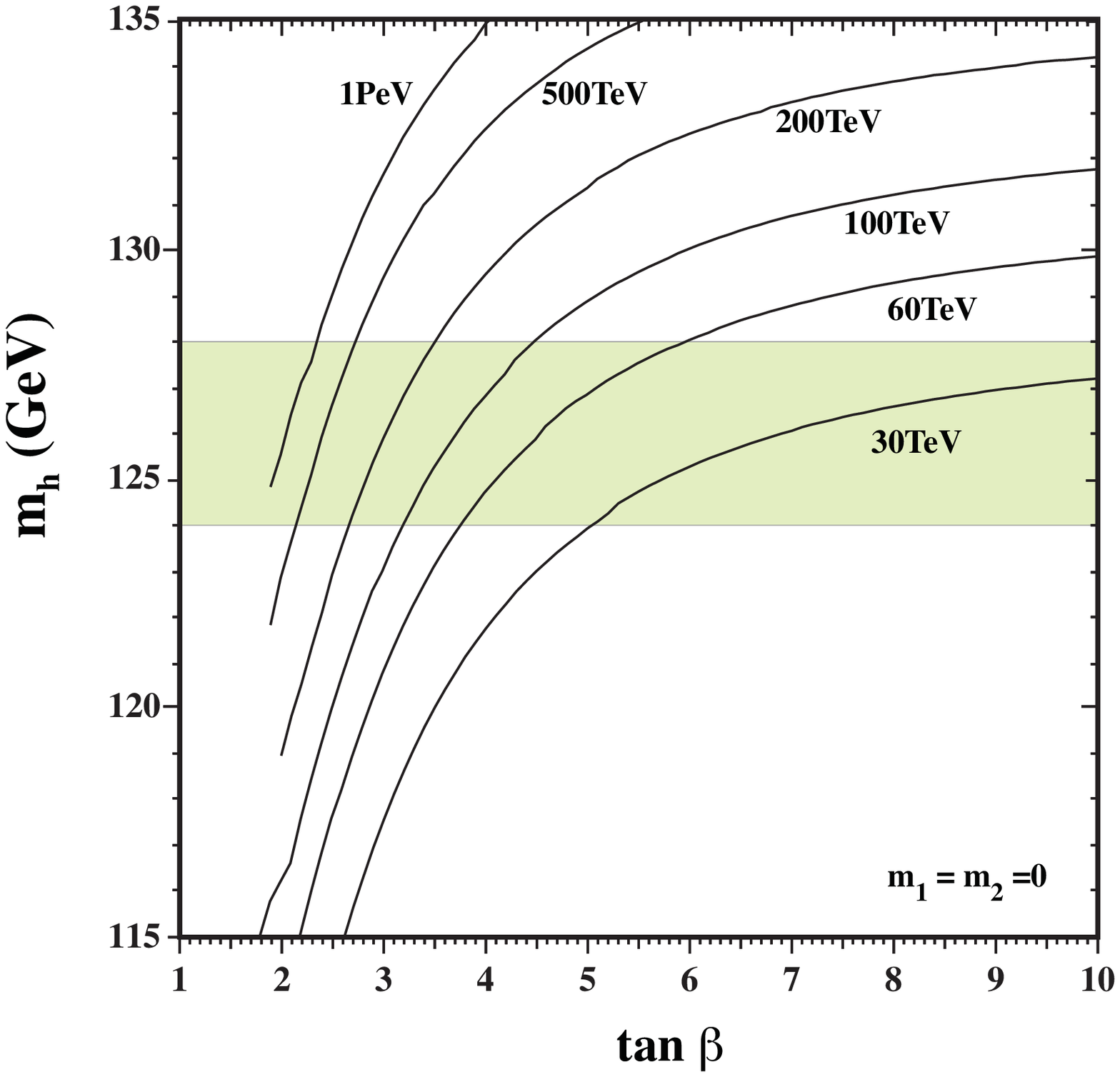,height=3.1in}
\epsfig{file=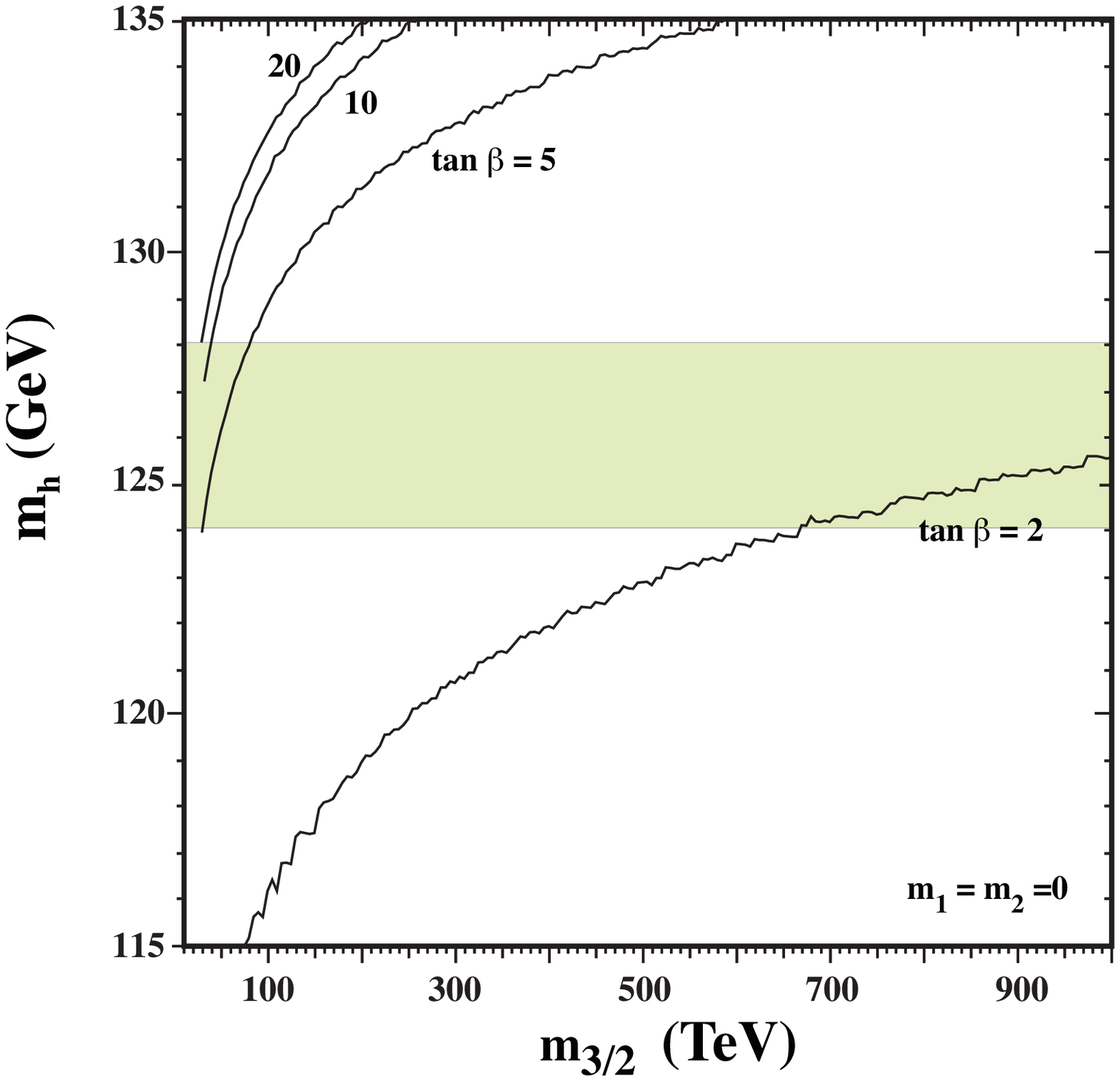,height=3.1in}
\hfill
\end{minipage}
\caption{
{\it
The light Higgs mass as a function of $\tan \beta$ (left) and $m_{3/2}$ (right) assuming
$m_1 = m_2 = 0$.
In the left panel, values of $m_h$ are shown for specific choices of $m_{3/2}$
from $30$--$1000$\,TeV as labeled.  In the right panel, $m_h$ is shown for $\tan \beta = 2, 5, 10$, and $20$.
For large values of $\tan \beta$,
we are restricted to relatively low values of $m_{3/2}$, near the lower bound of 30 TeV
imposed by the LEP limit on the chargino mass.
For low values of $\tan \beta$, the opposite is true and we are pushed to relatively
large $m_{3/2}$ as in the universal PGM model.
The LHC range (including an estimate of uncertainties) of $m_h \simeq 126 \pm 2$ GeV
is shown as the pale green horizontal band.
}}
\label{fig:nuhm2}
\end{figure}

The resulting Higgs masses as a function of $\tan \beta$ and $m_{3/2}$ are
shown in the left and right panels of Fig.\,\ref{fig:nuhm2} respectively.
In the left panel, we show the Higgs mass for specific choices of the gravitino mass
varying from $30$--$1000$\,TeV.  At large values of $\tan \beta$, the gravitino mass
must lie near its lower bound of 30 TeV imposed by the LEP limit on the chargino mass \cite{LEPsusy}.
At this value of $m_{3/2}$, the gluino mass is 1\,TeV, near its current LHC bound \cite{lhc}.
Such low values of $m_{3/2}$ are not possible in universal PGM models and
are realized here only because of the non-universality in the Higgs sector, which
facilitates the electroweak minimization conditions. In fact, in this scenario the only constraint on $\tan\beta$ comes from requiring that the Yukawa couplings remain perturbative.
At large $m_{3/2}$, we are forced to small $\tan \beta$ as in the universal model.
The same tendencies are seen in the right panel which shows
the Higgs mass as a function of $m_{3/2}$ for four fixed values of $\tan \beta = 2, 5, 10,$ and $20$.

\begin{figure}[ht]
\begin{minipage}{8in}
\epsfig{file=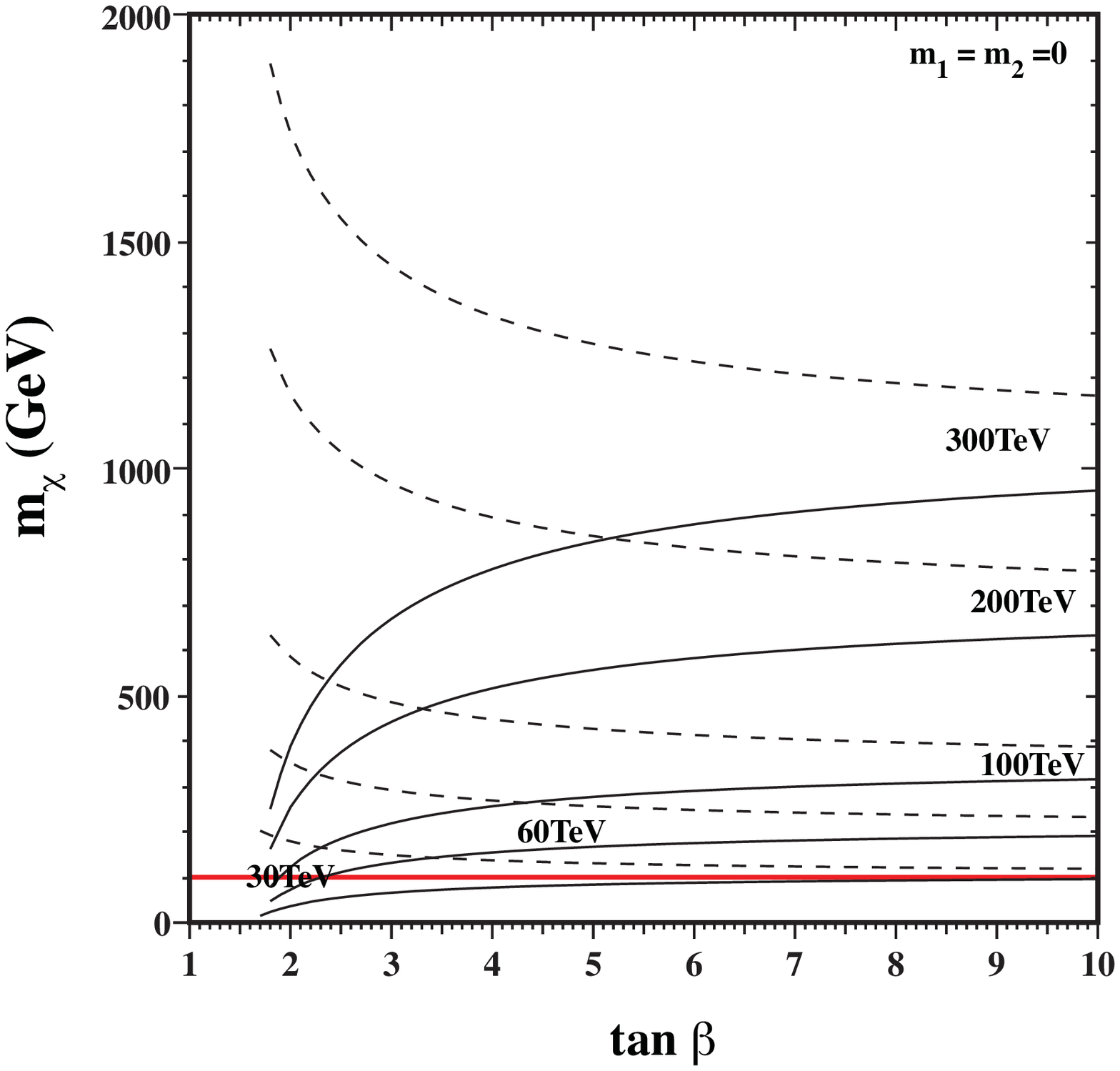,height=3.1in}
\epsfig{file=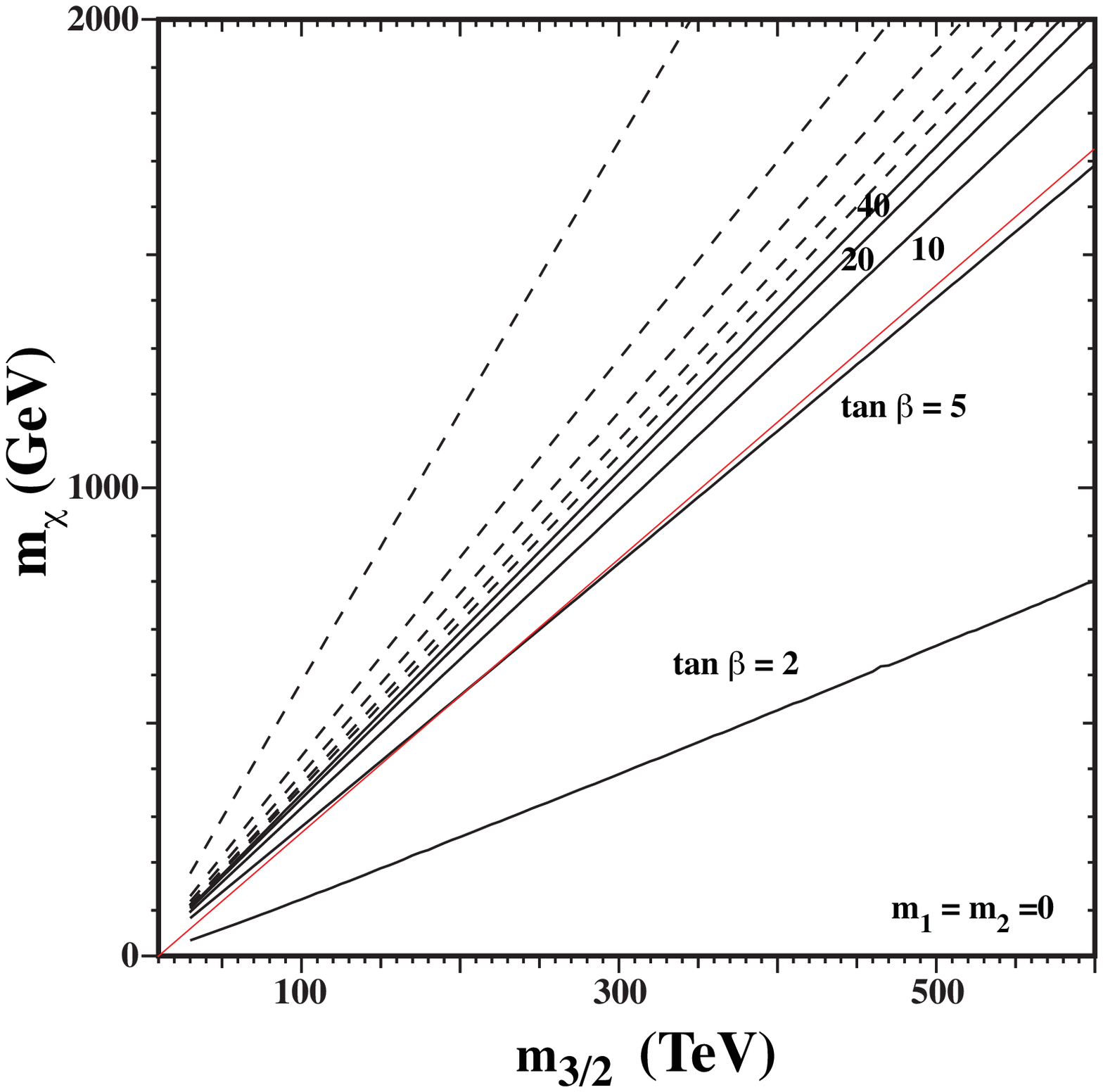,height=3.1in}
\hfill
\end{minipage}
\caption{
{\it
The wino mass as a function of $\tan \beta$ (left) and $m_{3/2}$ (right) for both
$\mu > 0$ (solid) and $\mu < 0$ (dashed).
For large values of $\tan \beta$, the differences between the wino mass at for $\mu > 0$ and $\mu < 0$
are less pronounced.
In the left panel, the LEP bound of $104$\,GeV on the chargino mass is shown as the horizontal red line.The different curves correspond to different values of $m_{3/2}$. Only the curves with $\mu < 0$
are labelled for clarity.
 In the right panel, the diagonal red line shows the wino mass when one-loop radiative corrections
 are ignored. Here the curves correspond to five values of $\tan \beta = 2, 5, 10, 20$, and $40$.
 }}
\label{fig:nuhm2mchi}
\end{figure}

For the specific case of $m_1 = m_2 = 0$, we also show the behavior of the neutralino
mass as a function of $\tan \beta$ and $m_{3/2}$  in the left and right panels of Fig.\,\ref{fig:nuhm2mchi} respectively. In the left panel, we see the $\tan\beta$ dependence which is predominantly driven by the corrections found in
Eq.\,(\ref{M2Thresh}). As $\tan\beta$ increases, the correction decreases and $m_\chi$ moves closer to its anomaly mediated contribution. In the right panel, we see that the neutralino mass is offset from the anomaly mediated contribution. This is due to threshold corrections of the sfermions. Since these corrections are not proportional to $\mu$, they are sub-leading to those found in Eq.\,(\ref{M2Thresh}) for small $\tan\beta$. However, in the large $\tan\beta$ limit they begin to dominate and give the offset seen in the right panel of Fig.\,\ref{fig:nuhm2mchi}.

\begin{figure}[h]
\vskip 0.5in
\vspace*{-0.45in}
\centering
\epsfig{file=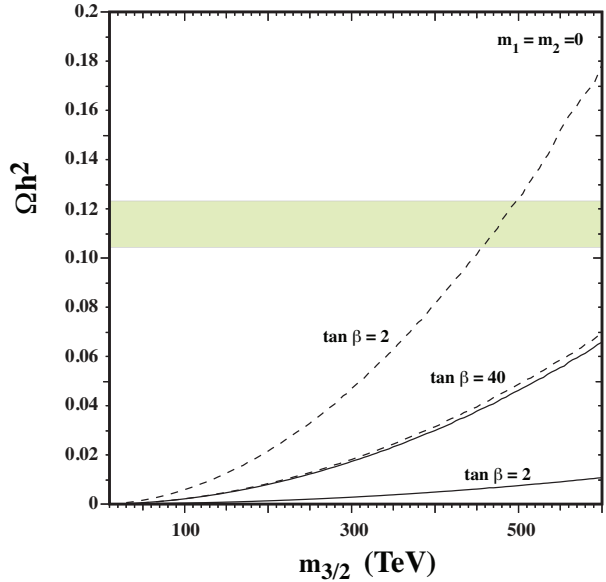,height=3.1in}
\caption{
{\it
The relic density, $\Omega_\chi h^2$
as a function of $m_{3/2}$ for both
$\mu > 0$ (solid) and $\mu < 0$ (negative).
 The WMAP range for the relic density is shown by the pale green horizontal band.
 The relic density crosses this band for $\mu <0$ when $m_{3/2} \simeq 460$--$500$\,TeV.
 At these gravitino masses the Higgs mass is in excess is about $125$\,GeV as seen in Fig.\,\protect{\ref{fig:nuhm2}}. This is reminiscent of the result in the universal PGM model.
}}
\label{fig:mgohsq}
\end{figure}

Finally, we show the behavior of the relic density in the case where $m_1 = m_2 = 0$. The variation in the relic density seen in Fig.\,\ref{fig:mgohsq} is due to variations in the mass of the wino.  As $m_{3/2}$ increases the wino mass increases and so does the relic density.  As $\tan\beta$ increases the wino mass increases (decreases) for $\mu < 0$ ($\mu > 0$) and so the relic density increases (decreases).  From this figure we see that only smaller values of $\tan\beta$ with $\mu<0$ will be able to give the correct thermal relic abundance and simultaneously predict a Higgs mass of $126.2 \pm 2$.
At lower $m_{3/2}$ or higher $\tan \beta$, we require either an alternative dark matter candidate
or rely on non-thermal mechanisms for neutralino production  \cite{ggw,hep-ph/9906527,Ibe:2004tg,Acharya:2008bk,dlmmo}.

\section{Summary}
Pure gravity mediation, with large tree level scalar masses and light anomaly mediated gauginos masses,
is one approach to getting a $126$ GeV lightest Higgs boson.
A particularly simple two parameter version of this model can be made consistent with all experimental constraints. The two free parameters can be chosen as $m_{3/2}$ and $c_H$, the coefficient of the generalized GM term.  The  coupling, $c_H$, can be traded for $\tan\beta$ making the phenomenology more transparent.
However, in these models, successful EWSB requires $\tan\beta \simeq 1.7$--$2.5$.
To get $m_h\simeq126\pm 2$, the gravitino mass then needs to be in the range $m_{3/2}\simeq 300$--$1500$ TeV. These large values for $m_{3/2}$ are the major drawback to this very simple model. If $m_{3/2}$ is this large, detection will be rather difficult because the gluino mass will be larger than about $6$\,TeV.
Though difficult to detect, these models are not without hope. Because the charged and neutral wino are nearly degenerate, the charginos will exhibit long lived charged tracks. Detection may be possible if the winos are light enough.

The heavy gluino of the universal case can be remedied if we take non-universal Higgs masses. With non-universal Higgs masses, the model can have up to $4$ parameters. Since the Higgs soft masses are now free parameters, we are allowed to take $m_2(M_{GUT})<m_{3/2}$. Since the boundary mass of $m_2^2$ is smaller than the stop squared masses, the top Yukawa coupling does not need to be so large in order to drive $m_2^2$ negative. This removes EWSB constraints on $\tan\beta$. The remaining constraints on $\tan\beta$ are from perturbativity of the Yukawa couplings. Larger values of $\tan\beta$ have two major effects. The tree level contribution to the lightest Higgs mass will be drastically increased. Because the tree level contribution is much larger, the loop corrections need to be smaller, leading to smaller stop masses.  Smaller stop masses require smaller $m_{3/2}$. In fact, in this scenario the most stringent constraint on $m_{3/2}$ is not from the lightest Higgs boson mass, as in the universal case, but is instead comes from the LEP bounds on the lightest chargino. The LEP bound on the lightest chargino gives $m_{3/2}>30$\,TeV, and the gluino mass can be of order a TeV and within reach of the LHC. The other important effect is to reduce the size of the threshold corrections to the wino mass. Since the the largest threshold correction is proportional to $-\mu$, this means the wino mass will increase (decrease) for $\mu >0$ ($\mu < 0$).  For very large values of $\tan\beta$, the wino mass approaches the value expected in anomaly mediation.

Interestingly, these non-universal models allow for a parameter space that is suggestive of the GUT or Planck scale theory. One interesting possibility, which we discussed here, is vanishing Higgs masses at the GUT scale.  Vanishing soft masses for the Higgs bosons can be realized in three different ways.
One possibility is a non-universal coupling to the supersymmetry-breaking field, $Z$, and
choosing the couplings $a,b$ appropriately.
A second possibility is a partial no-scale like structure where only the Higgs soft masses are zero.
This partial no-scale structure for the Higgs bosons can be broken by {a generalized GM term},
giving a linearly independent $\mu$ and $B\mu$ term which are necessary for EWSB, while still keeping the Higgs soft masses zero. The third possibility is that one or both of the Higgs bosons are Nambu-Goldstones of a broken symmetry. The gauge interactions of the MSSM, as well as the Yukawa and bilinear interactions of the superpotential, act as explicit breaking of this symmetry. This scenario is insightful as it may help explain why the Yukawa and gauge couplings are so small at the GUT or Planck scale. The phenomenology at the weak scale for these models is quite similar to more general non-universal Higgs models since $\tan\beta$ can still vary over the entire range allowed by perturbative Yukawa couplings.

\appendix

\section{No-Scale Higgs Masses\label{Append}}
Here, we calculate the mass terms for a fairly generic scalar potential reminiscent of our K\"ahler potential. For more details see \cite{Inoue:1991rk}. Starting with a K\"ahler potential,$G$,
we denote holomorphic (antiholomorphic) derivatives of $G$ with subscripts (superscripts).
The scalar potential is given by
\beq
V = e^G \left(G_i {G^{-1}}^i_j G^j - 3 \right).
\eeq
In this appendix we take $M_P=1$. The masses for the scalars are found to be
\begin{eqnarray}
\langle V_k^l\rangle = G_k^l(V+m_{3/2}^2)+ \mu_{ki}\mu^{il} -m_{3/2}^2G_i[G_{jk}^{il}-G^{il}_pG^{p}_{jk}]G^j\ ,
\end{eqnarray}
where we have taken $e^G=m_{3/2}^2$ and
\begin{eqnarray}
\mu_{ij} =\langle e^{G/2}( G_{ij} -G_\alpha G^\alpha_{ij}) \rangle \label{MuTe}\ ,
\end{eqnarray}
with the assumption that $\langle G_i^j \rangle =\delta _i^j$ for matter scalars. Our K\"ahler potential is of the form
\begin{eqnarray}
K=y_i\bar y_i-3\ln\left( 1- J(Z,\bar Z) - {\tilde J}(H_i,\bar H_i)\right)-3\ln\left(\rho+\bar\rho\right).
\end{eqnarray}
For this form of the K\"ahler potential we have the identity
\begin{eqnarray}
G_k^{\alpha i} G^k_{\beta j}-G^{\alpha i}_{\beta j} =-\frac{1}{3}\delta ^i_j \delta^\alpha_\beta\ ,
\end{eqnarray}
where Greek indices correspond to the hidden sector fields $Z$ and Latin indices correspond to the visible sector fields $H_i$.  This expression relies on the fact that the only non-zero first derivatives of $G$ are $G_\alpha,G^{\beta}$. For Latin indices corresponding to the $y_i$ fields, this expression vanishes. Using the above relation, we find
\begin{eqnarray}
\langle V_k^l\rangle = G_k^l(V+m_{3/2}^2)+ \mu_{ki}\mu^{il} -e^G\frac{1}{3}G_ZG^Z\delta_k^l\ ,
\end{eqnarray}
for the $H_i$ fields. Using now the vacuum condition, $\langle V \rangle = G_ZG^Z -3 =0$ we find
\begin{eqnarray}
\langle V_k^l\rangle =  \mu_{ki}\mu^{il}\ .
\end{eqnarray}
Since $\mu_{ij}$ is just the supersymmetric mass term, we see that the soft masses vanish. For the Higgs bosons, the K\"ahler potential is of the prescribed form and so the soft masses vanish. For the $y_i$ fields we have
\begin{eqnarray}
\langle V_k^l\rangle = m_{3/2}^2\delta_k^l+ \mu_{ki}\mu^{il},
\end{eqnarray}
and we see that the soft mass for the $y_i$ fields is $m_{3/2}$. If we add to the K\"ahler potential a piece of the form

\begin{eqnarray}
\Delta K = F(H_i)+ \bar F(\bar H_i)\ ,
\end{eqnarray}
 which would account for the Giudice-Masiero term outside of the logarithm, the above conclusions do not change because these additional terms do not contribute to any term with a holomorphic and anti-holomorphic derivative and so we still get $m_{H_i}^2=0$. If the Higgs bilinear term is inside the logarithm, the previous discussion readily applies and again we have $m_{H_i}^2=0$. This means we have the liberty to add the Giudice-Masiero term in or out of the natural logarithm with no change in the Higgs soft masses. The $B\mu$ relations, on the other hand, vary depending on the placement of the Giudice-Masiero term.  The $B\mu$-terms are found from
\begin{eqnarray}
V_{kl} &=& e^G\left[ G_{\alpha kl} G^\alpha +G_\beta (G^{-1})^\beta_{\alpha,kl} G^\alpha +2G_{kl} \right] ,
\end{eqnarray}
where we have neglected some unimportant terms.  Using our ansatz about the K\"ahler potential, this expression reduces to
\begin{eqnarray}
V_{kl} &=& m_{3/2}^2\left[-\frac{\mu'}{m_{3/2}M_P^2}G_\alpha G^\alpha-\Theta\frac{1}{3} \frac{c_H}{M_P^2}G_\alpha G^\alpha+2\left(c_H+\frac{\mu'}{m_{3/2}}\right) \right],
\end{eqnarray}
where each term above comes from the respective term in the preceding equation and in the case of strong moduli stabilization $\mu^\prime=\mu_0/(\rho+\bar\rho)^{3/2}$ where $\mu_0$ is the bilinear mass in the superpotential. $\Theta$ is either one or zero depending on whether
the Higgs bilinear term is in or out of the logarithm.  If the Higgs bilinear term is in the logarithm, $(G^{-1})^\beta_{\alpha}$ contains the Higgs bilinear  term and so $(G^{-1})^\beta_{\alpha,kl}$ contributes and $\Theta=1$.  If the Higgs bilinear term is outside of the logarithm, $(G^{-1})^\beta_{\alpha,kl}=0$ and $\Theta=0$. Simplifying we find
\begin{eqnarray}
B\mu= (2-\Theta)m_{3/2}^2c_H -m_{3/2}\mu' .
\end{eqnarray}

The $\mu$ term can be found from Eq. (\ref{MuTe}) which gives in both cases
\begin{eqnarray}
\mu=c_H m_{3/2} +\mu' .
\end{eqnarray}

If we now modify the above discussion, so that the K\"ahler potential is of the form
\begin{eqnarray}
K=-3\ln\left(\rho+\bar\rho+ J'(Z,\bar Z) + {\tilde J'}(H_i,\bar H_i)\right) ,
\end{eqnarray}
the calculation proceeds in roughly the same manner. However, a simple means to arrive at the correct answer is to factor out the $(\rho +\bar\rho)$ and redefine the Higgs fields. Doing this we find for the Higgs bilinear term outside of the logarithm,
\begin{eqnarray}
&&m_{H_i}^2=(\rho +\bar \rho)^2\left(c_H m_{3/2}+\mu'\right)^2=\mu^2, \\
&&B\mu= (\rho+\bar\rho) \left(m_{3/2}^2c_H-m_{3/2}\mu'\right),
\end{eqnarray}
which are identical to the previous case except $\mu$ and $B\mu$ are re-scaled by $(\rho +\bar \rho)$.
Now if the Higgs bilinear term is outside the logarithm, the $c_H$ term and $\mu$ term scale differently.  In this case we just re-scale $\mu$ by $(\rho +\bar \rho)$ and we get
\begin{eqnarray}
&&m_{H_i}^2= \left(2(\rho +\bar \rho)\mu' +m_{3/2}c_H\right)^2=\mu^2 ,\\
&&B\mu=m_{3/2}^2c_H-(\rho +\bar \rho) \mu' .
\end{eqnarray}

Lastly, we discuss the case where only one of the Higgs fields is inside of the logarithm. In this case one of the Higgs fields is the same as the $y_i$ fields and so they receive a soft mass just like the $y_i$ fields. However, they still have a supersymmetric mass giving a total boundary Higgs mass of
\begin{eqnarray}
m_{H_i}^2=m_{3/2}^2 +(\mu+c_Hm_{3/2})^2.
\end{eqnarray}
The $\mu$ and $B\mu$ relationships are identical to those for both the Higgs bilinear term in and out of the logarithm.
For the case with $\rho+\bar \rho$ also in the logarithm, all these relationships will be modified by factors of $\rho+\bar \rho$ which can be determined by the re-scaling arguments.
\section*{Acknowledgments}
The work of J.E. and K.A.O. was supported in part
by DOE grant DE--FG02--94ER--40823 at the University of Minnesota.
This work is also supported by Grant-in-Aid for Scientific research from the
Ministry of Education, Science, Sports, and Culture (MEXT), Japan, No.\ 22244021 (T.T.Y.),
No.\ 24740151 (M.I.), and also by the World Premier International Research Center Initiative (WPI Initiative), MEXT, Japan.

\end{document}